\documentclass[pdflatex,sn-mathphys-num, iicol]{sn-jnl}

\usepackage{graphicx}%
\usepackage{multirow}%
\usepackage{amsmath,amssymb,amsfonts}%
\usepackage{amsthm}%
\usepackage{mathrsfs}%
\usepackage[title]{appendix}%
\usepackage{xcolor}%
\usepackage{textcomp}%
\usepackage{manyfoot}%
\usepackage{booktabs}%
\usepackage{algorithm}%
\usepackage{algorithmicx}%
\usepackage{algpseudocode}%
\usepackage{listings}%
\usepackage{tabularx}
\usepackage{array}
\usepackage{subcaption}   

\usepackage{listings}
\usepackage{xcolor}
\usepackage{booktabs}
\usepackage{ragged2e}
\usepackage{xurl}

\lstdefinelanguage{proverif}{
  keywords={process, free, query, let, in, if, new, event, inj, insert, get, phase, table, data, type, attacker, out, fun, equation, set, select, yield, newchannel, channel, const, alias, and, or, not, true, false, else, reduc, forall},
  keywordstyle=\color{blue}\bfseries,
  morecomment=[l]{//},
  morecomment=[s]{(*}{*)}, 
  commentstyle=\color{gray}\itshape,
  morestring=[b]",
  stringstyle=\color{red},
  sensitive=true,
}

\lstdefinelanguage{tamarin}{
  keywords={
    theory, begin, end, 
    builtins, functions, equations, 
    rule, restriction, lemma, 
    diff, Fr, In, Out, 
    All, Ex, not, or, and
  },
  keywordstyle=\color{blue}\bfseries,
  morecomment=[l]{//},
  morecomment=[s]{/*}{*/},
  commentstyle=\color{gray}\itshape,
  morestring=[b]",
  sensitive=true,
  showstringspaces=false,
}

\theoremstyle{thmstyleone}%
\theoremstyle{thmstyletwo}%

\theoremstyle{thmstylethree}%

\begin{document}

\title[Article Title]{Bridging Theory and Practice: An Executable Taxonomy of Security Properties for ProVerif and Tamarin}


\author*[1]{\fnm{Leonard} \sur{Tudorache}}\email{l.s.tudorache@tue.nl}

\author[1]{\fnm{Ivan} \sur{Kurtev}}

\author[1]{\fnm{Mark} \sur{van den Brand}}

\affil*[1]{\orgdiv{Department of Mathematics and Computer Science}, \orgname{Eindhoven University of Technology}, \orgaddress{\street{Groene Loper 3}, \city{Eindhoven}, \postcode{5612 AE}, \state{North Brabant}, \country{The Netherlands}}}


\abstract{Security is critical for everything relying on modern digital systems. Because almost all digital interactions are governed by the Internet and cryptographic protocols, these protocols must serve as reliable mechanisms that guarantee core security properties, such as confidentiality and integrity. Formal verification of these protocols is a critical step in securing interconnected systems. Tools such as ProVerif and Tamarin are widely employed to perform automated verification. However, their effective use demands specialized domain knowledge, creating a significant learning curve for security protocol designers who often have a security, rather than a formal verification background. We therefore need structured, accessible resources to help protocol designers to express their design and requirements in the language of the formal verification tools.

To address this, we introduce a systematic and evidence-based taxonomy of security properties. This taxonomy is derived from a literature review of 53 recent studies (2022-2025) that used ProVerif and Tamarin, providing an up-to-date view of verified properties. We systematically categorize and define these properties, providing both informal definitions for intuitive comprehension and rigorous formal definitions expressed in first-order logic for clarity and consistency. We further detail modeling patterns and implement executable examples in both ProVerif and Tamarin, collected in an open repository. This work advances the state of the art by bridging the gap between theoretical security property definitions and their practical, executable verification models.}

\keywords{Formal Verification, Taxonomy, Security Properties, Modeling Patterns}



\maketitle

\section{Introduction}\label{sec:introduction}

Security protocols are used to define the way interconnected systems interact. They include a step-by-step description, usually accompanied by a sequence diagram that aims to guide the implementation. Furthermore, security properties are defined as requirements for such protocols, which can be verified by hand or in an automated manner before the actual implementation.

Formal verification techniques offer a structured approach to identifying flaws in security protocols, making them a crucial component in the design process. Numerous automated tools have been developed for this purpose, including ProVerif~\cite{blanchet2018proverif}, Tamarin~\cite{meier2013tamarin}, Scyther~\cite{cremers2008scyther}, AVISPA~\cite{AVISPA}, and Verifpal~\cite{10.1007/978-3-030-65277-7_8}, among others. These tools allow their users to model their security protocols and specify the security properties to be checked. Then they perform a verification process where all the possible paths are verified, providing step-by-step traces that highlight if a property does not hold.

Automated verification tools for security protocols often require domain-specific and tool-specific expertise to be used effectively, and each tool has different capabilities and limitations. As a result, users must acquire in-depth knowledge of multiple tools to thoroughly and formally verify the security of their protocols. Moreover, developers of the automated verification tools usually have a formal verification background and might have a different approach than protocol designers, who often have a security background. This mismatch in expertise often results in a steep learning curve, as protocol designers must gain additional formal verification knowledge to model and analyze their systems accurately. For instance, to verify a simple authentication property in Tamarin, a protocol designer must learn to express it as multiset rewriting rules over a global state, while the same property in ProVerif requires expressing it in terms of processes and events in ProVerif's applied pi calculus-based input language. These are two fundamentally different formalisms that demand significant background knowledge before any protocol-specific work can begin. Developers of automated verification tools need to invest in transferring observations back to the protocol world, ensuring that formal analysis results are meaningful and actionable for security engineers. Thus, there is a clear need for a user-friendly approach that abstracts the complexity involved in formally verifying security protocols.

As we highlighted in our previous work~\cite{292d45b0bcaf4dd3a75b2b2ba47bd853}, there is a lack of efficient and robust security protocols~\cite{tawalbeh_iot_2020}, and the complexity of security testing is high. The next step after defining a security protocol is to verify it. There are various approaches available to verify protocols, ranging from mathematical proofs to automated verification via specialized tools. Nevertheless, all require a deep understanding of the security properties targeted by a protocol. 
There are various classifications of security properties in the literature~\cite{Aldini_2006, Focardi_Gorrieri_Martinelli_2004, Sayar_Messe_Ebersold_Bruel_2025}, but these do not include all security properties that are verified nowadays and often use a lower level of abstraction, by using rigorous formal specification to define the properties, making it more difficult for security protocol designers to understand them. Furthermore, they do not provide examples of how to model them and integrate them into security protocols.

The lack of a systematic classification creates a gap in understanding how security properties are used in security protocols by practitioners and how these can be modeled and verified using automated verification tools. Therefore, there is a need for a structured taxonomy that defines security properties that are currently verified by protocol designers. To the best of our knowledge, there is no classification of security properties that provides a common and formal understanding of all security properties identified in the current literature.
Therefore, a taxonomy, which is a structured system for classifying and organizing concepts into groups (often hierarchical)~\cite{Nickerson_Varshney_Muntermann_2013}, would provide a foundation for security protocol verification, to help protocol designers understand how to model fundamental security protocols and verify them using automated verification tools. Furthermore, this taxonomy, paired with modeling patterns, gives users a clear path from property selection to executable verification, cutting through the varying definitions and tool-specific syntax that currently make the process unnecessarily difficult.

We propose a taxonomy by systematically extracting and categorizing security properties verified with ProVerif and Tamarin over the past three years from the tool-based property verification literature. We aim to provide both formal and informal definitions and create a repository containing example models for ProVerif and Tamarin that represent how each property of the taxonomy can be specified and verified. This aims to aid protocol designers in starting their verification process using these tools. First, the taxonomy provides both an informal and a formal definition of the security property. Then, the users can employ the modeling patterns presented in the ProVerif and Tamarin models for each security property, following a simple guide on how to apply the verification queries to their protocols, or use the example models as a starting skeleton for a protocol.

In this paper, we focus on the automated verification tools ProVerif and Tamarin, two prominent automated approaches dedicated to analyzing security protocols. Both use an unbounded number of sessions and the Dolev-Yao model~\cite{dolev_security_1983}. The Dolev-Yao model is an abstract adversarial framework employed in the formal verification of security protocols, which simplifies analysis by assuming that cryptographic primitives are perfect while modeling an adversary with comprehensive control over all network communication channels. 

ProVerif operates by translating the models of security protocols into Horn clauses to analyze violations of the specified properties~\cite{blanchet2018proverif}. Conversely, Tamarin employs a state-based approach utilizing multiset rewriting rules~\cite{meier2013tamarin}. The protocol is modeled as a set of rewrite rules that modify a global multiset of facts, which represents the system's current state and event history. This approach enables stateful reasoning and the application of temporal logic (such as LTL) to specify and verify complex security properties

Tamarin provides native support for complex algebraic properties such as Diffie-Hellman exponentiation~\cite{diffie2022new}, full support for mutual global state, and stateful reasoning with temporal logic, enabling verification of properties such as post-compromise security. However, these capabilities come at the cost of undecidable verification problems~\cite{Belfaik_Lotfi_Sadqi_Safi_2024}. ProVerif, by contrast, offers more limited support for equational reasoning~\cite{Cortier_Grimm_Lallemand_Maffei_2017} and global state, but provides built-in biprocesses for modeling and verifying indistinguishability, and generally benefits from more automated and decidable verification.

The remainder of the article is structured into six sections. Section \ref{sec:related-work} provides a critical review of existing classifications and literature pertaining to security property formalization and verification techniques. Section \ref{sec:methodology} describes the goal, research questions, and defines the search protocol employed to identify and select relevant primary studies from the literature. Subsequently, Section \ref{sec:results} answers the research questions by describing a taxonomy of security properties, a comprehensive quantitative analysis of the selected papers, and a description of the modeling patterns. Section \ref{sec:case-study} aims to validate the applicability of the taxonomy and modeling patterns. Section \ref{sec:discussion} offers an interpretation and critical analysis of these findings, discussing the advancements over prior work and addressing threats to validity. Finally, Section \ref{sec:conclusion} summarizes the main contributions and addresses future work. 

\section{Related Work}\label{sec:related-work}

Although there is a common agreement on the definitions of security properties for security protocols, they are still open to interpretation since some of the formal definitions have not been widely agreed upon \cite{Focardi_Gorrieri_Martinelli_2004}. Focardi et al. \cite{Focardi_Gorrieri_Martinelli_2004} claim that a formal model is required to define the problem, and then the formal definition of the security property needs to be presented concerning the model. In this regard, they propose a classification of security properties by formulating it through a general scheme, Generalized Non Deducibility Composition (GNCD), to allow for an easier formal comparison of formal security properties. They use CryptoSPA~\cite{10.5555/647544.730459} to specify security properties, but this does not allow for automated formal verification. 

Rouland et al.~\cite{Rouland_Hamid_Bodeveix_Filali_2019} specify security properties of a system using a technology-independent specification language: first-order logic. In the next phase, they specify the designed system using Alloy~\cite{10.5555/2141100}, a specific computational model of the system used for model checking and verification. The security properties specified are: confidentiality, integrity, and availability. On top of these security properties, Sayar et al.~\cite{Sayar_Messe_Ebersold_Bruel_2025} also include authorization, authenticity, and accountability as being known as properties that a system should preserve. They propose a System Development Life Cycle (SDLC) taxonomy that allows experts to identify threats and propose defenses by synchronizing security properties with attacks, defenses, and system assets. Event-B~\cite{10.5555/1855020} language was used to define the taxonomy, evaluate it, and update it as it evolves. 

Aldini~\cite{Aldini_2006} proposes a classification of security properties using process algebra based on Linda-like coordination primitives~\cite{10.1145/2363.2433} (a shared, associative memory where data items can be added), where they used relaxed notions of behavioral equivalences that include the external observer's abilities. The results highlight the positive influence of Linda coordination model on the expressivity. Hermann et al.~\cite{hermann_taxonomy_2025} propose a taxonomy of functional implementation-level security features. This taxonomy is achieved by conducting a systematic literature review, then mapping the results to security standards and producing a relation to famous security frameworks. This aims to help developers select appropriate security features to achieve certain standards. 

In contrast to prior taxonomies or classifications, which have either remained at a conceptual level~\cite{Focardi_Gorrieri_Martinelli_2004, Sayar_Messe_Ebersold_Bruel_2025} or targeted functional security features~\cite{hermann_taxonomy_2025}, our work advances the state of the art in three ways. First, we systematically extract and categorize security properties verified with ProVerif and Tamarin over the past three years, providing an up-to-date and evidence-based view. Second, we propose informal and formal definitions, expressed in first-order logic, to ensure clarity and consistency. Third, we make the taxonomy executable by mapping each property to representative models in both ProVerif and Tamarin, collected in an open repository. This combination of rigor, reproducibility, and practical applicability distinguishes our work from earlier classifications.

\section{Methodology}\label{sec:methodology}

This paper aims to provide a taxonomy of the current security properties verified using automated verification tools such as ProVerif and Tamarin over the past three years. To find relevant papers, we have defined a methodology in Section \ref{sec:protocol} based on established guidelines for systematic literature reviews (SLRs) presented by Kitchenham et al. \cite{KITCHENHAM20097-guidelines-slr}. This facilitates the identification and analysis of resources used to address specific research questions in a manner that is both reproducible and unbiased. This section aims to present the research questions and the protocol used to extract relevant studies from the literature.

\subsection{Goal and Research Questions}
The primary objective of this study is to advance the understanding of how security properties are modeled and verified in formal protocol analysis using ProVerif and Tamarin, two of the most widely adopted tools in the field. To achieve this, we define the following research questions:

\begin{itemize}
  \item [RQ1] \textit{What security properties are verified using ProVerif and Tamarin?} This question seeks to compile and categorize the most frequently verified security properties in protocol verification literature that utilizes ProVerif and Tamarin. The goal is to understand which properties (e.g., secrecy, authentication, unlinkability) are prioritized by researchers, and how often they appear across case studies and domains.

  \item [RQ2] \textit{How are security properties formally and informally defined?} This question addresses the dual nature of defining security properties: (i) informal definitions, often presented in natural language to convey intuitive understanding, and (ii) formal definitions, expressed in logical formalisms (e.g., first-order logic) to support rigorous verification. The objective is to assess the clarity, precision, and consistency of these definitions across studies.
  
  \item [RQ3] \textit{What patterns and best practices can be identified for modeling and verifying common security properties of protocols using ProVerif and Tamarin?} This question aims to uncover recurring modeling structures, abstraction choices, and verification strategies used across different studies that employ ProVerif and Tamarin for protocol analysis. The focus is on identifying methodological patterns and tool-specific best practices that enhance the correctness, scalability, and interpretability of security property verification.

\end{itemize}

\label{sec:protocol}
\subsection{Protocol}

We define a protocol that includes the search string, strategy, inclusion/exclusion criteria, and data extraction to enhance the reproducibility of our study. We performed a literature search on the following digital libraries: ACM Digital Library, IEEE Xplore, ScienceDirect, Scopus, and WebOfScience. The search string, comprising ("security" OR "cryptographic") AND "protocol" AND "verification" AND ("tamarin" OR "proverif"), was executed on the title, abstract, and keyword fields. Furthermore, we applied a filter on the years the papers were published, from January 2022 to April 2025. By focusing on this period, this work aims to provide a contemporary perspective on the security properties currently verified by protocol designers, also including properties from previous security property classification papers. This resulted in 454 papers (after removing duplicates). Table \ref{tab:search_protocol} defines a search protocol that includes the search string, inclusion/exclusion criteria, and the data extraction form. The inclusion/exclusion criteria state that the selected papers must provide a ProVerif or Tamarin model, a description of the security properties, and be published in the past three years. After applying the protocol, we selected a total of 53 papers that met the requirements. Next, data extraction is performed to extract the ProVerif or Tamarin models along with the definition of the security properties from the selected papers.

\begin{table*}[ht]
\centering
\caption{Protocol overview}
\begin{tabularx}{\textwidth}{lX}
\toprule
\textbf{Search string} &
("security" OR "cryptographic") AND "protocol" AND "verification" AND ("tamarin" OR "proverif") \\
\addlinespace

\textbf{Search strategy} &
Database search: Scopus, IEEE Xplore, Web of Science, ACM Digital Library, SpringerLink. \newline
Forward and backward snowballing using Google Scholar. \\
\addlinespace

\textbf{Inclusion and exclusion criteria} &
\textbf{Inclusion:} \newline
Articles, conference papers, book chapters, and workshop papers. \newline
Studies where the title, abstract, or keywords suggest that their main topic of study is \textbf{security protocol verification using ProVerif or Tamarin}. \newline
Studies published between January 2022 and April 2025. \newline
Articles that are electronically accessible. \newline
\newline
\textbf{Exclusion:} \newline
Articles not in English. \newline
Studies such as slides, websites, or conference reviews. \newline
Articles that do not include a model. \newline
Articles that are not electronically accessible. \\
\addlinespace

\textbf{Data extraction} &
Title, year, security properties (formal/informal definitions), tool used (ProVerif or Tamarin), model \\
\bottomrule
\end{tabularx}
\label{tab:search_protocol}
\end{table*}

The next step is to extract a relevant security protocol model for each security property in both ProVerif and Tamarin. The models are obtained by either extracting a specific model from previous studies, tools' manuals, or by deriving it from the selected papers. These models aim to be used as a starting point for modeling protocols that aim to maintain these security properties.

\section{Results}\label{sec:results}

This section presents the results of our study by addressing each research question. First, we provide an overview of the security properties identified in the literature by proposing a taxonomy. Second, we provide informal and formal definitions in first-order logic for each property. Finally, we describe some of the modeling patterns identified from the extracted models.

\subsection{RQ1 - What security properties are verified using ProVerif and Tamarin?}
\label{sec:results-rq1}

The initial extraction of security properties from the selected papers yielded a total of 64 distinct terms. However, several of these overlapped semantically or represented the same concept expressed differently across studies. We performed a normalization process to ensure terminological consistency, grounding our classification in the functional goal of each property. By focusing on these functional outcomes, hierarchical or protocol-specific terms were consolidated under a common representative (see Table \ref{tab:normalization}). For instance, \textit{conditional privacy-preserving} and \textit{ticket privacy} were both grouped under Privacy, and \textit{authenticity} and \textit{message integrity} were both unified under the Integrity group. Furthermore, we have identified protocol-specific properties (such as fairness, eligibility, or verification of intent) that, based on our analysis, are mechanisms that involve multiple properties already identified. We define a primitive security property as an atomic, self-contained security requirement that cannot be decomposed into simpler ones, serving as a building block from which more complex security mechanisms can be constructed. Therefore, the scope of our taxonomy is primitive security properties.

\begin{table*}[ht]
\centering
\caption{Normalization of extracted security properties into the final taxonomy}
\label{tab:normalization}
\begin{tabularx}{\textwidth}{|X|X|}
\hline
\textbf{Original Property (extracted from selected papers)} & \textbf{Mapped / Final Category} \\
\hline
explicit key confirmation, strong secrecy & Confidentiality – Secrecy \\
perfect forward security & Confidentiality – Forward secrecy \\
data confidentiality & Confidentiality \\
anonymous, identity anonymity, user anonymity, identity protection, identity privacy & Privacy – Anonymity \\
conditional privacy-preserving, ticket privacy & Privacy – Privacy \\
indistinguishability, single-blindness, long-term unlinkable & Privacy – Unlinkability and untraceability \\
message integrity, authenticity & Integrity – Integrity \\
non-equivocation, unforgeability & Integrity – Subproperties \\
trace, traceable, traceability & Accountability – Traceability \\
non-repudiation & Accountability – Non-repudiation \\
transferable authentication, strong user authentication & Authentication – Authentication \\
fairness, eligibility, verification of intent, correctness, profile binding, receipt-freeness, no-double-spending, vote verifiability, offline verifiable, remote attestation & Protocol-specific (excluded) \\
\hline
\end{tabularx}
\end{table*}

\begin{figure}
    \centering
    \includegraphics[width=\linewidth]{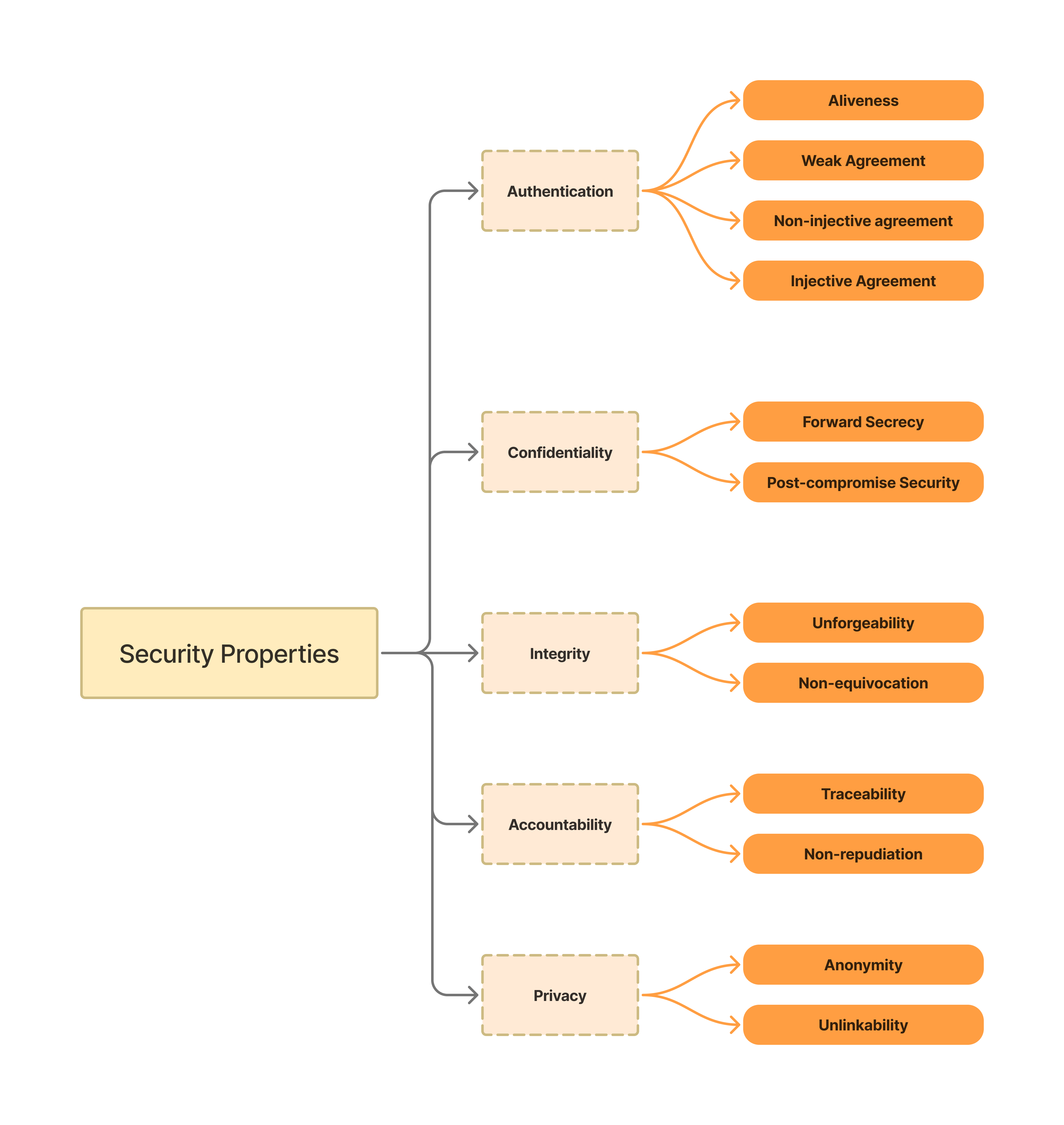}
    \caption{Taxonomy of Security Properties}
    \label{fig:taxonomy-diagram}
\end{figure}

Figure \ref{fig:taxonomy-diagram} presents the taxonomy of security properties extracted from the selected papers.  We define five main categories: authentication, integrity, confidentiality, privacy, and accountability. Each category represents a high-level security property that can be further extended to achieve a stricter security level. For instance, for confidentiality, ensuring key secrecy provides a certain level of confidentiality, but checking for forward secrecy ensures that if an attacker obtains a key, they cannot decrypt messages sent previously.

To further investigate RQ1, a quantitative analysis was conducted to systematically examine the occurrence, distribution, and relationships of security properties. The analysis aims to enable an evidence-based understanding of how ProVerif and Tamarin are applied in practice. Data was extracted into a structured dataset, normalized for consistency, and analyzed.

\begin{figure*}
    \centering
    \includegraphics[width=0.9\linewidth]{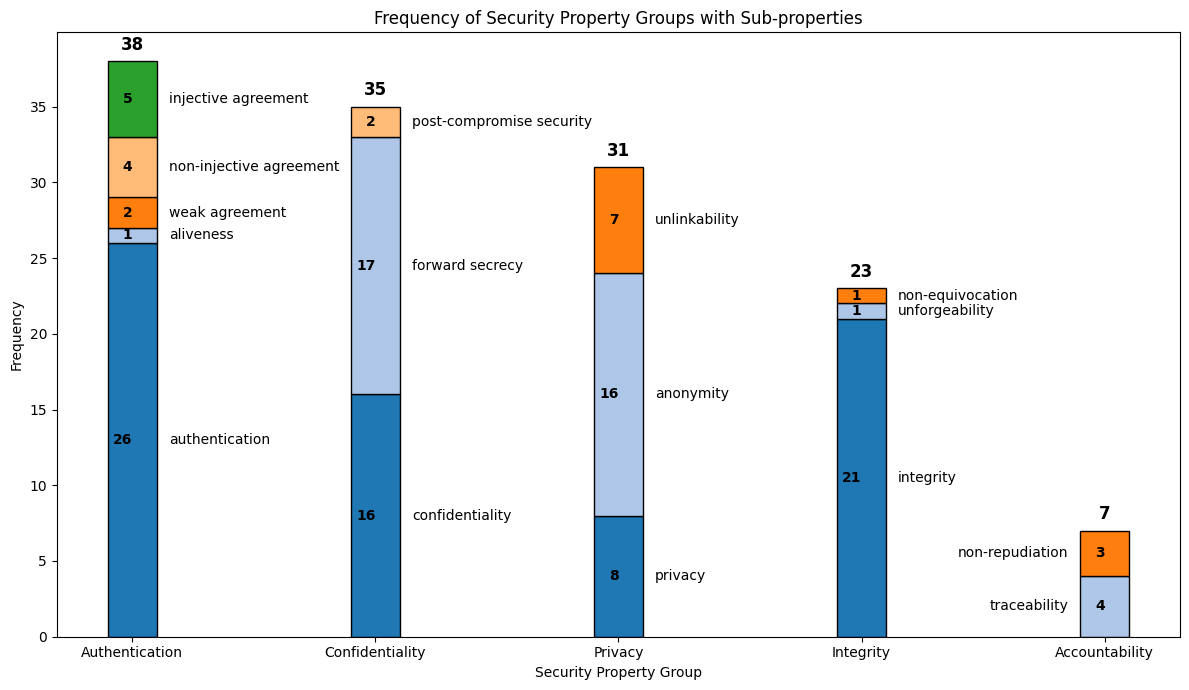}
    \caption{Frequency of Security Properties per Group}
    \label{fig:frequency-properties}
\end{figure*}

Figure \ref{fig:frequency-properties} illustrates the distribution of the verified security properties by each group defined in our taxonomy. The most verified property is authentication (38 occurrences), but confidentiality (35), privacy (31), and integrity (23) are not far from it. This shows that the current verification tools provide the necessary means to verify these properties. Nevertheless, accountability (7 occurrences) is the least verified property from the selected papers. The reason behind this might be that to verify accountability, the users need to model more than the message flows, they need to model evidence generation and evaluation by a judge or a third party. Most protocol verification focuses on preventing attacks (such as ensuring confidentiality and integrity) rather than on proving responsibility after an incident. Additionally, many case studies that have been identified in the literature are based on well-known protocols (TLS, Signal), where accountability is not a primary design goal. The distribution of the sub-categories shows that core, broadly applicable security properties, such as integrity, forward secrecy, confidentiality, and anonymity dominate security protocol verification studies. This suggests that automated verification efforts predominantly focus on widely applicable and well-supported properties, such as secrecy, privacy, and authentication, while more specialized or complex properties remain underexplored. This might indicate that properties that require more effort to be modeled and verified are less likely to be verified. Table \ref{tab:taxonomy_mapping} maps each selected paper to the security properties defined by our taxonomy, where each property is grouped by its high-level category.

\begin{figure}[ht]
    \centering
    \includegraphics[width=0.8\linewidth]{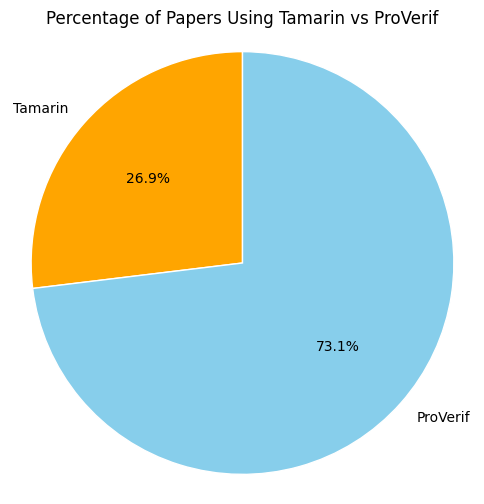}
    \caption{Distribution of papers by verification tool.}
    \label{fig:distribution-verification-tool}
\end{figure}

\begin{figure}[ht]
    \centering
    \includegraphics[width=\linewidth]{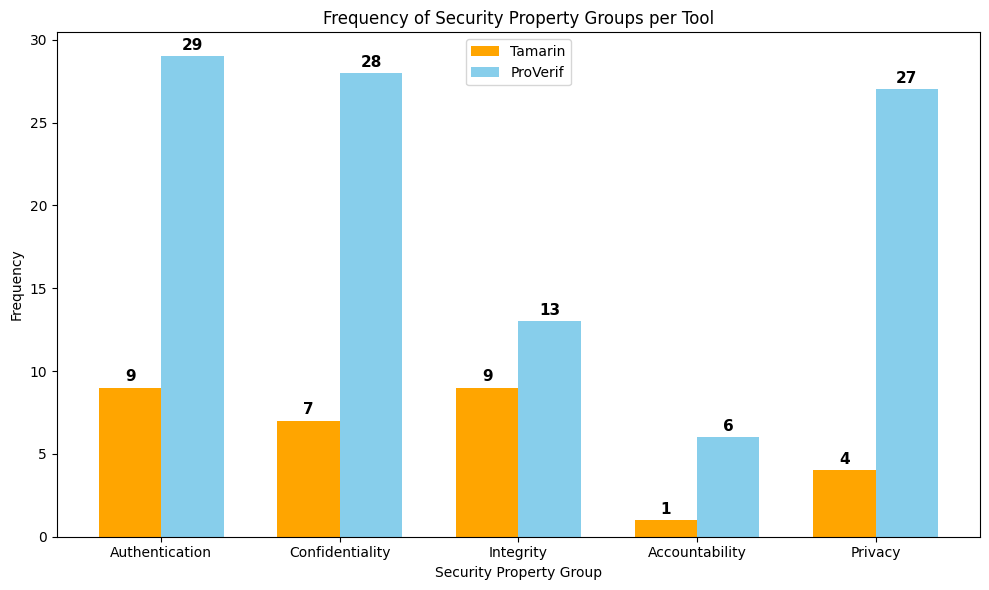}
    \caption{Frequency of properties verified per tool.}
    \label{fig:freq-property-per-tool}
\end{figure}

Figure \ref{fig:freq-property-per-tool} presents the distribution of security properties per each taxonomy group by the verification tool used. This highlights the verification tool that is predominantly used to verify certain properties. Taking into consideration that only 26.9\% of the selected papers use Tamarin (see Figure \ref{fig:distribution-verification-tool}), the proportion for which both Tamarin and ProVerif are used for authentication, confidentiality, and integrity is broadly similar. However, for accountability and privacy, Tamarin tends to be used less compared to ProVerif. ProVerif’s higher automation, long‑standing support for equivalence‑based privacy, and extensive documentation explain why it is used more often in practice, while Tamarin is favored when one needs more expressive trace‑based reasoning at the cost of higher modeling and proof effort. Thus, protocol designers have a lower learning barrier when selecting ProVerif to model and verify their protocol.

Both ProVerif and Tamarin are used to verify most of the security properties, although each tool may be more straightforward when it comes to modeling certain properties. Tamarin is used to verify post-compromise security properties, while ProVerif is used to check traceability. For instance, Tamarin is used for post-compromise security because it can explicitly model mutable state, temporal reasoning, and compromise events. In contrast, traceability, which belongs to the accountability class of properties, is often expressed as a reachability condition. This is more straightforwardly verified in ProVerif using its event-based correspondence queries. In general, the properties defined in our taxonomy can be verified by both Tamarin and ProVerif, but they may differ in the level of detail with which the protocols are modeled. Furthermore, the modeling approach might differ per tool since each tool offers different capabilities.

\begin{table*}[htbp]
\centering
\caption{Mapping of selected papers to security properties verified}
\label{tab:taxonomy_mapping}
\begin{tabular}{|l|l|p{9cm}|}
\hline
\textbf{Category} & \textbf{Property} & \textbf{Papers} \\ \hline
Authentication & Authentication / Mutual Auth. & \cite{Jacomme_Klein_Kremer_Racouchot_2023}, \cite{X.Ren_J.Cao_B.Niu_L.Gan_Y.Zhang_L.Xiong_Y.Luo_H.Li_2024}, \cite{Wang_Li_Guan_2024}, \cite{Chunka_Banerjee_SachinKumar_2023}, \cite{Méré_Jouault_Pallardy_Perdriau_2024}, \cite{H.Feng_J.Guan_H.Li_X.Pan_Z.Zhao_2023}, \cite{S.Bussa_R.Sisto_F.Valenza_2023b}, \cite{Moustafa_Sethi_Aura_2025}, \cite{DeVaere_Stoger_Perrig_Tsudik_2024}, \cite{M.Pradhan_S.Mohanty_2024}, \cite{Y.Huang_G.Xu_X.Song_Y.Xu_2024}, \cite{Ahn_Kwak_Kim_2024}, \cite{Saini_Kaur_Kumar_Kumar_2024}, \cite{Zou_Cao_Lu_Wang_Xu_Ma_Cheng_Xi_2024}, \cite{Liu_Zhou_Cao_Xu_Wang_Gao_Zeng_Xu_2023}, \cite{You_Kim_Pawana_Ko_2024}, \cite{Ko_Pawana_Won_Astillo_You_2024}, \cite{Ahmed_Peltonen_Sethi_Aura_2024}, \cite{Duttagupta_Marin_Singelee_Preneel_2023}, \cite{Wu_Meng_Yang_Kumari_Pirouz_2022}, \cite{Xie_Wu_Hou_2023}, \cite{Zarbi_Zaeembashi_Bagheri_Adeli_2024}, \cite{Kim_Ryu_Lee_Lee_Won_2023}, \cite{Li_2023}, \cite{Miculan_Vitacolonna_2023} \\ \hline
Authentication & aliveness & \cite{Akman_Ginzboorg_Damir_Niemi_2023} \\ \hline
Authentication & weak agreement & \cite{Akman_Ginzboorg_Damir_Niemi_2023}, \cite{X.Ren_J.Cao_B.Niu_L.Gan_Y.Zhang_L.Xiong_Y.Luo_H.Li_2024} \\ \hline
Authentication & non-injective agreement & \cite{Akman_Ginzboorg_Damir_Niemi_2023}, \cite{X.Ren_J.Cao_B.Niu_L.Gan_Y.Zhang_L.Xiong_Y.Luo_H.Li_2024}, \cite{Bodei_DeVincenzi_Matteucci_2024}, \cite{Wang_Laing_Moreira_Ryan_2024} \\ \hline
Authentication & injective agreement & \cite{Akman_Ginzboorg_Damir_Niemi_2023}, \cite{X.Ren_J.Cao_B.Niu_L.Gan_Y.Zhang_L.Xiong_Y.Luo_H.Li_2024}, \cite{Bodei_DeVincenzi_Matteucci_2024}, \cite{Bursuc_Horne_Mauw_Yurkov_2023}, \cite{Zhu_Xu_Cui_2024} \\ \hline
Confidentiality & & \cite{Ahmed_Peltonen_Sethi_Aura_2024}, \cite{X.Ren_J.Cao_B.Niu_L.Gan_Y.Zhang_L.Xiong_Y.Luo_H.Li_2024}, \cite{Wang_Li_Guan_2024}, \cite{Xie_Wu_Hou_2023}, \cite{Feng_Wang_Bobda_2025}, \cite{H.Feng_J.Guan_H.Li_X.Pan_Z.Zhao_2023}, \cite{Zhu_Xu_Cui_2024}, \cite{S.Bussa_R.Sisto_F.Valenza_2023a}, \cite{Watanabe_Yoneyama_2024}, \cite{S.Bussa_R.Sisto_F.Valenza_2023b}, \cite{Fujita_Yoneyama_2024}, \cite{Ahn_Kwak_Kim_2024}, \cite{Ko_Pawana_Won_Astillo_You_2024}, \cite{Wang_Wu_Wen_Zhou_Hu_Xie_2025}, \cite{Seo_Kim_Lee_Kwon_Seo_2023}, \cite{Méré_Jouault_Pallardy_Perdriau_2024} \\ \hline
Confidentiality & forward secrecy & \cite{Akman_Ginzboorg_Damir_Niemi_2023}, \cite{Miculan_Vitacolonna_2023}, \cite{Zou_Cao_Lu_Wang_Xu_Ma_Cheng_Xi_2024}, \cite{Liu_Zhou_Cao_Xu_Wang_Gao_Zeng_Xu_2023}, \cite{Rangwani_Om_2023}, \cite{Li_2023}, \cite{Chunka_Banerjee_SachinKumar_2023}, \cite{Wu_Meng_Yang_Kumari_Pirouz_2022}, \cite{Y.Huang_G.Xu_X.Song_Y.Xu_2024}, \cite{Wang_Wu_Wen_Zhou_Hu_Xie_2025}, \cite{Cremers_Jacomme_Naska_2023}, \cite{You_Kim_Pawana_Ko_2024}, \cite{Zarbi_Zaeembashi_Bagheri_Adeli_2024}, \cite{Xie_Liu_Ding_Tan_Han_2023}, \cite{Ko_Pawana_Won_Astillo_You_2024}, \cite{Kim_Ryu_Lee_Lee_Won_2023}, \cite{Jacomme_Klein_Kremer_Racouchot_2023} \\ \hline
Confidentiality & post-compromise security & \cite{Kim_Ryu_Lee_Lee_Won_2023}, \cite{Cremers_Jacomme_Naska_2023} \\ \hline
Integrity & & \cite{Ahmed_Peltonen_Sethi_Aura_2024}, \cite{Miculan_Vitacolonna_2023}, \cite{Shahrouz_Analoui_2024}, \cite{Raimondo_Bernardi_Marrone_Merseguer_2023}, \cite{Y.Huang_G.Xu_X.Song_Y.Xu_2024}, \cite{Wang_Wu_Wen_Zhou_Hu_Xie_2025}, \cite{Sabry_Samavi_2022}, \cite{Xie_Wu_Hou_2023}, \cite{Feng_Wang_Bobda_2025}, \cite{Wagner_Birnstill_Beyerer_2024}, \cite{Seo_Kim_Lee_Kwon_Seo_2023}, \cite{Baloglu_Bursuc_Mauw_Pang_2024}, \cite{Watanabe_Yoneyama_2024}, \cite{Fujita_Yoneyama_2024}, \cite{Hu_Zhang_Weimerskirch_Mao_2022}, \cite{Pietro_Savio_Roberto_2023}, \cite{Ahn_Kwak_Kim_2024}, \cite{Y.Song_F.Jiang_S.W.AliShah_R.Doss_2023}, \cite{B.Fila_S.Radomirović_2024}, \cite{S.Lu_Z.Li_X.Miao_Q.Han_J.Zheng_2023}, \cite{Ko_Pawana_Won_Astillo_You_2024} \\ \hline
Integrity & unforgeability & \cite{Lafourcade_Mahmoud_Marcadet_Olivier-Anclin_2024} \\ \hline
Integrity & non-equivocation & \cite{Giantsidi_Pritzi_Gust_Katsarakis_Koshiba_Bhatotia_2025} \\ \hline
Accountability & traceability & \cite{Shahrouz_Analoui_2024}, \cite{Xie_Wu_Hou_2023}, \cite{Li_Jin_Levchenko_2024}, \cite{Duttagupta_Marin_Singelee_Preneel_2023} \\ \hline
Accountability & non-repudiation & \cite{Shahrouz_Analoui_2024}, \cite{H.Feng_J.Guan_H.Li_X.Pan_Z.Zhao_2023}, \cite{Jacomme_Klein_Kremer_Racouchot_2023} \\ \hline
Privacy & & \cite{Shahrouz_Analoui_2024}, \cite{Xie_Wu_Hou_2023}, \cite{H.Feng_J.Guan_H.Li_X.Pan_Z.Zhao_2023}, \cite{Zhu_Xu_Cui_2024}, \cite{Fujita_Yoneyama_2024}, \cite{S.Lu_Z.Li_X.Miao_Q.Han_J.Zheng_2023}, \cite{Wang_Laing_Moreira_Ryan_2024}, \cite{Lafourcade_Mahmoud_Marcadet_Olivier-Anclin_2024} \\ \hline
Privacy & anonymity & \cite{Shahrouz_Analoui_2024}, \cite{Y.Huang_G.Xu_X.Song_Y.Xu_2024}, \cite{Saini_Kaur_Kumar_Kumar_2024}, \cite{Wang_Wu_Wen_Zhou_Hu_Xie_2025}, \cite{Xie_Wu_Hou_2023}, \cite{Kim_Ryu_Lee_Lee_Won_2023}, \cite{S.Bussa_R.Sisto_F.Valenza_2023a}, \cite{Xie_Liu_Ding_Tan_Han_2023}, \cite{S.Lu_Z.Li_X.Miao_Q.Han_J.Zheng_2023}, \cite{Rakeei_Giustolisi_Lenzini_2023}, \cite{Lafourcade_Mahmoud_Marcadet_Olivier-Anclin_2024}, \cite{Zou_Cao_Lu_Wang_Xu_Ma_Cheng_Xi_2024}, \cite{Liu_Zhou_Cao_Xu_Wang_Gao_Zeng_Xu_2023}, \cite{Rangwani_Om_2023}, \cite{Li_2023}, \cite{Jacomme_Klein_Kremer_Racouchot_2023} \\ \hline
Privacy & unlinkability & \cite{X.Ren_J.Cao_B.Niu_L.Gan_Y.Zhang_L.Xiong_Y.Luo_H.Li_2024}, \cite{Xie_Wu_Hou_2023}, \cite{Li_Jin_Levchenko_2024}, \cite{H.Feng_J.Guan_H.Li_X.Pan_Z.Zhao_2023}, \cite{S.Bussa_R.Sisto_F.Valenza_2023a}, \cite{Xie_Liu_Ding_Tan_Han_2023}, \cite{Bursuc_Horne_Mauw_Yurkov_2023} \\ \hline
\end{tabular}
\end{table*}

\subsection{RQ2 - How are security properties formally and informally defined?}
\label{sec:results-rq2}

This section aims to present a taxonomy of primitive security properties that are verified using automated formal verification tools such as ProVerif and Tamarin. These properties have been identified from the literature, mostly based on the papers published in the past three years. Security has been studied for decades, evolving from early concerns over physical and communication security to the rigorous study of information security in the digital era. We define a taxonomy of security properties (see Section \ref{sec:results-rq1}) for automated formal verification and in this section, we provide the formal definitions expressed in first-order logic. 

In first-order logic, predicate symbols are the building blocks used to express properties and relationships between entities. A predicate symbol takes one or more terms as arguments and evaluates to either true or false, essentially making a statement about those terms. For example, a predicate like \textit{Honest(A)} asserts that agent A is honest, while a binary predicate like \textit{Knows(A, k)} asserts that agent A knows key k. In the context of security protocol verification, predicates are used to capture protocol-specific concepts such as whether a message was sent, whether a session key was established, or whether an adversary has learned a secret. This is aligned with the way properties are specified in ProVerif and Tamarin, where events (in the case of ProVerif) or actions (in the case of Tamarin) form a trace during the execution of the protocol.
For our definitions, we first give an informal definition, then introduce the predicates, and finally, introduce the formal definition. 

The proposed definitions have been adapted from established formalizations in the literature to ensure consistency and to provide a robust foundational layer for modeling, enabling the practical verification of security properties through automated analysis tools. Furthermore, we provide examples on how to model and verify these security properties using both Tamarin and ProVerif. We first model the properties using sequence diagrams (as a tool-independent notation) that describe step by step the interaction between actors of the protocols. Next, we develop ProVerif and Tamarin models for each property based on the models presented in the selected papers (see Section \ref{sec:results-rq3}).

\subsubsection{Authentication}
The authentication definitions are based on the authentication levels defined by Lowe~\cite{Lowe_1997}: aliveness, weak agreement, non-injective agreement, and injective agreement. Particularly, for authentication, we consider two agents, A and B, taking the roles of initiator and responder, respectively. We aim to ensure that each agent can verify that it genuinely interacted with the other, that is, that the entity on the other end of the exchange is who it claims to be.
The formalization of the authentication is based on Lowe's definition and Tamarin's manual \cite{tamarinmanual}, where they are described in first-order logic.

\paragraph{Aliveness}

\subparagraph{Informal Definition}
This is the weakest form of authentication, which ensures that agent B has been running the protocol at some point in the past, but it does not guarantee that agent B interacted with agent A, or that the protocol has been running recently.

\subparagraph{Formal Definition}
\begin{itemize}
    \item $end(A, B, i)$ Agent A has completed a run of the protocol, apparently with agent B, at time (or trace step) i.
    \item $start(B, j)$ Agent B has started (or participated in) a protocol session at time (or trace step) j.
\end{itemize}

$\forall A, B, i:
end(A, B, i)
\implies \exists j: start(B,j)$

\paragraph{Weak Agreement}

\subparagraph{Informal Definition}
An initiator A has a weak agreement with another agent B, whenever A completes a run of the protocol, believing they interacted with B. Nevertheless, the fact that B was executing the protocol under the assumption of interacting with A does not necessarily imply that B assumed the role of responder.
\subparagraph{Formal Definition}
\begin{itemize}
    \item $end(A, B, t, i)$ Agent A has completed a run of the protocol, apparently with agent B, with term t (data shared among the participants), at time (or trace step) i.
    \item $start(B, A, t, j)$ Agent B has started (or participated in) a protocol session with agent A, with term t, at time j.
\end{itemize}


\begin{align*}
\forall\, A, B, t_1, i:\;
&\mathit{end}(A, B, t_1, i) \\
\implies\; &\exists\, t_2, j:\; \mathit{start}(B, A, t_2, j)
\end{align*}

\paragraph{Non-injective Agreement}

\subparagraph{Informal Definition}
An initiator A has a non-injective agreement with a responder B if A finishes the protocol believing they communicated with B, and B also has been running the protocol before, believing they were interacting with A, and both used the same data. However, A could run multiple instances of the protocol while B may not take part in all of them. Thus, it does not guarantee a one-to-one relationship.
\subparagraph{Formal Definition}

\begin{itemize}
    \item $end(A, B, t, i)$ Agent \textit{A} has completed a run of the protocol, apparently with agent \textit{B}, with term t, at time (or trace step) i.
    \item $start(B, A, t, j)$ Agent \textit{B} has started (or participated in) a protocol session with agent \textit{A}, with term \textit{t}, at time \textit{j}.
\end{itemize}

\begin{align*}
\forall\, A, B, t, i:\;
&\mathit{end}(A, B, t, i) \\
\implies\; &\exists\, j:\; \mathit{start}(B, A, t, j)
\end{align*}

\paragraph{Injective Agreement}

\subparagraph{Informal Definition}
This is the strongest level of authentication, where there is a non-injective agreement between initiator A and responder B with a strict one-to-one relationship. An initiator A has a non-injective agreement with a responder B if A finishes the protocol believing they communicated to B, and B also has been running the protocol before, believing they were interacting with A, and both used the same data. Additionally, every time agent A completes a run of the protocol, agent B has also been running the same instance of the protocol.
\subparagraph{Formal Definition}

\begin{itemize}
    \item $end(A, B, t, i)$ Agent \textit{A} has completed a run of the protocol, apparently with agent \textit{B}, with term \textit{t}, at time (or trace step) \textit{i}.
    \item $start(B, A, t, j)$ Agent \textit{B} has started (or participated in) a protocol session with agent \textit{A}, with term \textit{t}, at time \textit{j}.
\end{itemize}

\begin{align*}
\forall\, A, B, t, i:\;
&\mathit{end}(A, B, t, i) \\
\implies\; &\exists\, j:\; \mathit{start}(B, A, t, j) \wedge j < i \\
\wedge\; &\neg \exists\, A_2, B_2, i_2:\; \mathit{end}(A_2, B_2, t, i_2) \wedge i_2 \neq i
\end{align*}

\subsubsection{Integrity}

The goals of integrity are defined by Biba~\cite{Biba_1977} as data consistency is maintained both internally and externally, the data cannot be modified by unauthorized parties, where authorized refers to both cases where the party needs to be authorized to access a system, and it has the required levels of access to perform data modification. Furthermore, we include two more levels of integrity: unforgeability and non-equivocation.

\paragraph{Unforgeability}

\subparagraph{Informal Definition}
Unforgeability ensures that no adversary can produce a valid message or signature that was not generated by an honest participant, thus, an attacker cannot modify messages~\cite{Ahn_Kwak_Kim_2024, 10.1007/3-540-44706-7_20}.

\subparagraph{Formal Definition}

\begin{itemize}
    \item $valid_{sig}(A, m, sig)$ The signature \textit{sig} is valid for message m under agent \textit{A}’s public key
    \item $sign_{event}(A, m, sig, i)$ At time \textit{i}, agent \textit{A} actually signed message \textit{m}, producing signature \textit{sig}
\end{itemize}

\begin{align*}
\forall\, A, m, sig:\;
&\mathit{valid}_{sig}(A, m, sig) \\
\implies\; &\exists\, i:\; \mathit{sign}_{event}(A, m, sig, i)
\end{align*}

\paragraph{Non-equivocation}

\subparagraph{Informal Definition}
Equivocation happens when an actor of the protocol sends different messages to different actors in the same instance of the protocol while the protocol instructs that the same message must be sent~\cite{Clement_Junqueira_Kate_Rodrigues_2012}. Therefore, non-equivocation is defined as assurance that an actor cannot use the same key to validate contradicting messages. Alternatively, non-equivocation is defined as the guarantee that a single message cannot be accepted more than once~\cite{Giantsidi_Pritzi_Gust_Katsarakis_Koshiba_Bhatotia_2025}. 

\subparagraph{Formal Definition}
\begin{itemize}
    \item $valid_{sig}(k, m)$ Message \textit{m} is valid under public key \textit{k} — i.e., \textit{m} was signed with the private key corresponding to \textit{k}, and the signature verification succeeded.
\end{itemize}

\begin{align*}
\forall\, k, m_1, m_2:\;
&\mathit{valid}_{sig}(k, m_1) \\
\wedge\; &\mathit{valid}_{sig}(k, m_2) \\
\implies\; &m_1 = m_2
\end{align*}

\subsubsection{Confidentiality}
Confidentiality, also referred to as secrecy in practice, is achieved when a protocol ensures the secrecy of data M by never publishing M, or the means to compute M, even if there is an interaction with an attacker \cite{Abadi}. This also holds when the protocol is using public channels for communication. We include stronger versions of confidentiality, such as forward secrecy and post-compromise security. Confidentiality is one of the most verified security properties using ProVerif or Tamarin. ProVerif supports it as a primitive query to check this property, while Tamarin provides clear instructions in the manual~\cite{tamarinmanual}. Based on the lemma extracted from the Tamarin manual, we can express confidentiality as follows:
\begin{itemize}
    \item $K(x)$ Value \textit{x} is known by the attacker
    \item $secret(x)$ Value \textit{x} is expected to remain confidential — i.e., not known to the adversary.
\end{itemize}

$\forall x: secret(x) \implies \neg K(x)$

\paragraph{Forward Secrecy}

\subparagraph{Informal Definition}
Forward secrecy property, also known as perfect forward secrecy \cite{Krawczyk2019}, is a property of key exchange protocols that ensures past session keys remain secure even if long-term secret keys (used for authentication or key negotiation) are later compromised~\cite{Krawczyk_2021}.

\subparagraph{Formal Definition}

\begin{itemize}
    \item $sessionKey(k, t)$ The session key \textit{k} was generated or used in a protocol session at time \textit{t}.
    \item $longTermKey(ltk)$ \textit{ltk} is a long-term secret key, used for authentication or key agreement.
    \item $K(x)$ Value \textit{x} is known by the attacker
    \item $leaked(k, t)$ Key \textit{k} is leaked at time \textit{t}
\end{itemize}

\begin{align*}
\forall\, ltk, k_1, t_1, t_2:\; &\mathit{sessionKey}(k_1, t_1) \\
\wedge\; &\mathit{longTermKey}(ltk) \\
\wedge\; &\mathit{leaked}(ltk, t_2) \wedge t_1 < t_2 \\
\implies\; &\neg K(k_1)
\end{align*}

\paragraph{Post-compromise Security (PCS)}

\subparagraph{Informal Definition}
Post-compromise security, also known as backward secrecy~\cite{10.1145/3372297.3423354}, is defined as the property of a protocol where actor A has a security guarantee about the communication with actor B, even if the attacker has B's secrets~\cite{Cremers_Jacomme_Naska}. It has the ability to recover after the compromise, so leaking the partner's key does not imply that all future messages can be decrypted~\cite{Cremers_Jacomme_Naska_2023}. The recovery process is often a protocol-specific function, so it differs from protocol to protocol.

\subparagraph{Formal Definition}
Adapted from \cite{Cremers_Jacomme_Naska_2023}

\begin{itemize}
    \item $sent(sid_A, A, B, sck, i)$ Agent \textit{A} sends a message to \textit{B} using session key \textit{sck} during the session identified by $sid_A$, at time \textit{i}.
    \item $K(x, j)$ Value \textit{x} is known by the attacker at time \textit{j}.
    \item $heal(sid_A, A, B, k)$ The session between \textit{A} and \textit{B} (session id $sid_A$) was healed at time \textit{k}.
    \item $compromise(sid_A, A, B, l)$ The session between \textit{A} and \textit{B} is compromised at time \textit{l}.
\end{itemize}

\begin{align*}
\forall\, A, B, sck, sid_A, i, j, k:\; 
&\mathit{sent}(sid_A, A, B, sck, i) \\
\wedge\; &K(sck, j) \\
\wedge\; &\mathit{heal}(sid_A, A, B, k) \\
\wedge\; &k < i \\
\implies\; &\exists\, l:\; \mathit{compromise}(sid_A, A, B, l) \\
\wedge\; & k < l
\end{align*}

\subsubsection{Privacy}

According to Westin~\cite{Westin2015PrivacyAndFreedom}, privacy is defined as ‘The claim of individuals, groups, or institutions to determine for themselves when, how and to what extent information about them is communicated to others’. We further include four sub-categories of privacy in our taxonomy: anonymity and unlinkability.

\paragraph{Anonymity}

\subparagraph{Informal Definition}
Anonymity is defined as the confidentiality of the identification of agents of a protocol~\cite{Schneider_Sidiropoulos_1996, Samfat_Molva_Asokan_1995}. Only authorized parties have access to relevant identification elements.
\subparagraph{Formal Definition}

\begin{itemize}
    \item $actor(B, id)$ Agent \textit{B} is a legitimate actor in the protocol with unique identifier \textit{id}.
    \item $K(x)$ Value \textit{x} is known by the attacker
\end{itemize}

$\forall B, id: actor(B,id) \implies \neg K(id)$

\paragraph{Unlinkability}

\subparagraph{Informal Definition}
Unlinkability, also referred to as indistinguishability, represents the impossibility for an outside observer to distinguish two processes/systems~\cite{blanchet2018proverif, Baelde_Delaune_Moreau_2020}.

Furthermore, indistinguishability is one of the most important elements for ProVerif and Tamarin when it comes to the verification phase. If two processes are observationally equivalent, it means that no adversary, no matter how powerful, can tell which one they are interacting with. Thus, we present the following process notation, which describes that two processes are indistinguishable:\\

\subparagraph{Formal Definition}
$P(s) | P(s) \approx P(s) | P(t)$
\\

The notation expresses that two parallel compositions of process P are observationally equivalent, where on the left-hand side, both instances run with the same value \textit{s}, while on the right-hand side, one instance runs with \textit{s} and the other with a different value \textit{t}. The equivalence relation states that no external observer can distinguish between these two configurations based on the observable outputs of the process.

\subsubsection{Accountability}
Accountability is a property of a system or protocol that ensures that if a security goal is violated, the responsible parties can be identified~\cite{8823726}. The verification of this property usually requires a trusted third-party judge.

Based on the definition introduced by Künnemann et al.~\cite{8823726}:
\begin{itemize}
    \item $\mathit{Judged}(t,c,i)$ -- the judge
          received and accepted a response for certificate $c$ 
          carrying timestamp $t$, at time $i$
    \item $\mathit{Secret}(c,\mathit{sk}, j)$ -- the secret key $\mathit{sk}$ 
          is bound to certificate $c$, established at time $j$
    \item $K(\mathit{sk}, k)$ -- Value $\mathit{sk}$ is known by the attacker
          at time $k$
    \item $\mathit{Time}(t, l)$ -- the honest clock process emitted timestamp 
          $t$ at time $l$
\end{itemize}

\begin{align*}
\neg \exists\, c,\, \mathit{sk},\, t,\, i,\, j,\, k,\, l.\quad
&\mathit{Judged}(t,c,i) \\
\wedge\; &\mathit{Secret}(c,\mathit{sk},j) \\
\wedge\; &K(\mathit{sk},k) \\
\wedge\; &\mathit{Time}(t,l) \\
\wedge\; &k < l
\end{align*}

The intuition here is that if the secret \textit{sk} becomes known to the attacker, the certification process must have happened \textit{before} that.
The formula is a negation of the opposite situation: secret \textit{sk} is revealed before the time of the certification.

\paragraph{Traceability}

\subparagraph{Informal Definition}
Traceability is defined as the ability for a trusted entity to trace a message back to the sender to avoid malicious behaviors~\cite{Xie_Wu_Hou_2023, Fu_Qin_Wang_Li_Zhang_2017}.
\subparagraph{Formal Definition}

\begin{itemize}
    \item $send(A, m)$ Agent $A$ sends message $m$.
    \item $computeId(m)$ A function that returns the identity of the agent who actually sent message $m$.
\end{itemize}

$\forall A,m:  send(A, m) \implies computeId(m) = A$

\paragraph{Non-repudiation}

\subparagraph{Informal Definition}
Non-repudiation property for protocols ensures that its agents are able to provide evidence that another agent previously transmitted a message, which can be presented to a judge to prove its behavior~\cite{Schneider_1998}.
\subparagraph{Formal Definition}

\begin{itemize}
    \item $obs(m)$ Message \textit{m} is observed (an attacker or a participant of the protocol sees it on the network).
    \item $sent(m)$ Message \textit{m} was sent by a legitimate process.
\end{itemize}

$\forall m: obs(m) \implies sent(m)$\\

These informal and formal definitions provide input for the verification queries used by both ProVerif and Tamarin. Besides giving a thorough understanding of how each security property is verified, this taxonomy explicitly identifies the specific events required for protocol instrumentation and the logical assertions necessary for tool-specific queries.

\subsection{RQ3 - What patterns and best practices can be identified for modeling and verifying common security properties of protocols using ProVerif and Tamarin?}
\label{sec:results-rq3}

To answer RQ3, we identify the best practices and patterns used to develop both ProVerif and Tamarin models for the security properties presented in our taxonomy. We analyzed the models proposed in the selected papers. Then, we proposed a model for each security property in both ProVerif and Tamarin. These are small models that capture the essential behavior of a property. In the case that some properties are not specifically modeled in the selected papers, or they were only modeled using one tool, we perform a search in the literature or check examples from ProVerif or Tamarin manuals. This allows us to extract/build example models for those security properties. This aims to help security protocol designers start the modeling and verification process for their protocol. We created a repository~\cite{leonardtudorache_2026_19352740} with all the modeling patterns and security properties definitions extracted from the selected papers. Furthermore, our proposed models can be found in folders \textit{ProVerif Models} and \textit{Tamarin Models} in our repository.

\begin{table*}[htbp]
    \centering
    \small
    \renewcommand{\arraystretch}{1.4}
    \caption{Comparative Synthesis of Security Properties}
    \label{tab:security_synthesis}
    
    \begin{tabularx}{\textwidth}{@{} 
        >{\RaggedRight\hsize=0.9\hsize}X 
        >{\RaggedRight\hsize=0.9\hsize}X 
        >{\RaggedRight\hsize=1.0\hsize}X 
        >{\RaggedRight\hsize=1.2\hsize}X 
        >{\RaggedRight\hsize=1.0\hsize}X 
    @{}}
        \toprule
        \textbf{Property} & \textbf{Conceptual Maturity} & \textbf{Tool Support \newline (ProVerif vs. Tamarin)} & \textbf{Modeling Requirements} & \textbf{Verification Difficulty} \\
        \midrule
        
        \textbf{Confidentiality} & 
        \textbf{High} \newline Standard definitions (e.g., Secrecy, Forward Secrecy) are established. & 
        \textbf{ProVerif:} Native primitive queries. \newline 
        \textbf{Tamarin:} Strong support via lemmas. & 
        \textbf{Reachability:} \newline Requires defining secret terms and checking adversary derivability. & 
        \textbf{Low to Medium} \newline Standard secrecy is efficient. PCS is difficult due to state explosion. \\
        \addlinespace
        
        \textbf{Authentication} & 
        \textbf{High} \newline Based on Lowe's hierarchy (Aliveness to Injective Agreement). & 
        \textbf{ProVerif:} Excellent (Correspondence queries). \newline 
        \textbf{Tamarin:} Excellent (Logic formulas). & 
        \textbf{Correspondence:} \newline Requires annotating protocols with events (e.g., \texttt{begin} and \texttt{end}) to prove relationships. & 
        \textbf{Low} \newline Straightforward unless complex multi-party interactions are involved. \\
        \addlinespace
        
        \textbf{Integrity} & 
        \textbf{High} \newline Based on Biba and standard signature validity. & 
        \textbf{Both:} Supported via event matching (checking signatures/hashes). & 
        \textbf{Event Logic:} \newline Explicit modeling of signing and verification steps (e.g., \texttt{signed\_by}). & 
        \textbf{Medium} \newline Requires correct modeling of cryptographic primitives to prevent trivial violations. \\
        \addlinespace
        
        \textbf{Privacy} & 
        \textbf{Medium} \newline Well-defined (Unlinkability) but complex implementation. & 
        \textbf{ProVerif:} Uses \texttt{biprocesses} / \texttt{choice}. \newline 
        \textbf{Tamarin:} Uses Diff-equivalence. & 
        \textbf{Observational Equivalence:} \newline Modeling two distinct systems to prove they are indistinguishable to an attacker. & 
        \textbf{High} \newline Computationally heavy. Equivalence checks occur at every path, often causing non-termination. \\
        \addlinespace
        
        \textbf{Accountability} & 
        \textbf{Low} \newline Least verified; lacks standardized patterns. & 
        \textbf{ProVerif:} Good for Traceability. \newline 
        \textbf{Tamarin:} Better for Non-repudiation. & 
        \textbf{Structural:} \newline Requires modeling Evidence generation and third-party verifiers, not just message flows. & 
        \textbf{Very High} \newline High manual modeling effort; difficult to abstract into simple queries. \\
        
        \bottomrule
    \end{tabularx}
\end{table*}

Table \ref{tab:security_synthesis} presents a comparison of security properties based on ProVerif and Tamarin in terms of conceptual maturity, tool support, modeling requirements, and verification difficulty. There are three levels of conceptual maturity (low, medium, and high) based on how well-defined and standardized the property is within the current literature. These are based on the patterns and definitions identified in the selected papers and the tools' manuals. Furthermore, the tool support classification is extracted from the specific strengths and weaknesses of the two automated tools.  Modeling requirements describe the specific coding structures or architectural elements a user must implement to verify the property. Verification difficulty estimates the computational costs and the likelihood of a tool to converge based on our hands-on experience. 

For the authentication property, we have chosen the Needham-Schroeder-Lowe Public Key protocol from the Tamarin example models~\cite{Basin_Tamarin_Prover}. This model has been modified to abstract away from traditional adversarial key-compromise scenarios to prioritize a formal deconstruction of the Lowe hierarchy of authentication. By omitting the long-term key-reveal rules, the model establishes a \textit{closed-world} environment that isolates the protocol’s structural logic from the noise of forward secrecy analysis. This approach is motivated by the need to verify the incremental establishment of trust between actors, transitioning from basic aliveness to full injective agreement. Through the introduction of explicit state facts and lemmas for each authentication level, the model serves as a specialized investigative framework to demonstrate how we can verify the authentication levels using Tamarin.

Although we provide ProVerif and Tamarin models along with sequence diagrams for each property from the taxonomy in our repository, we select two properties to present in detail in this section. We choose unlinkability and unforgeability, two properties that are not explicitly described in either ProVerif or Tamarin manuals. 

As defined in Section \ref{sec:results-rq2}, unlinkability is defined as the indistinguishability of two processes/systems. This is a trivial property used for automated formal verification of security properties. In our case, we will limit the scope to the possibility of an attacker distinguishing between two processes of an actor of a protocol. Listing \ref{listing:pv-unlinkability} presents the ProVerif model where we check if an attacker can observe the difference between two sessions that use different session IDs. We use the construct \textit{choice} that allows for checking observational equivalence, such that an attacker cannot distinguish between two processes. Our model is just an example of how observational equivalence can be used to check unlinkability. This construct can also be used inside processes to check for the equivalence of output messages. However, if it is used inside a process, it will be computationally heavy as the check will take place at every path used for the verification, leading to non-termination for the verification of mid-level complexity. 
For Tamarin, we have a similar construct (\textit{diff}) for verifying observational equivalence that was used in this case inside the Tamarin rule on line 26 in Listing \ref{listing:tamarin-unlinkability}. In this case, Tamarin seems to handle better than ProVerif the non-termination problem for observational equivalence as the verification converges (when used in the equivalent of the ProVerif process).
We do not provide a sequence diagram for this model because it has one process/actor that sends only one message on the public channel, so there is no interaction between multiple parties. Nevertheless, the attacker can initiate a message exchange. 

\begin{lstlisting}[language=proverif, caption=Unlinkability Check in ProVerif, label=listing:pv-unlinkability]
free c : channel.
type uid. type nonce.
fun pid(uid, nonce): bitstring.

free U, V : uid [private].

let session(u: uid) =
  new r: nonce;
  (* out(c, u). *)
  out(c, pid(u, r)).

(* Biprocess: left = both sessions use U; right = second uses V *)
process
  ( session(U) | session(choice[U, V]) )
\end{lstlisting}

\begin{lstlisting}[language=tamarin, caption=Unlinkability Check in Tamarin, label=listing:tamarin-unlinkability]
// tamarin-prover --diff --prove Unlinkability.spthy

theory IdentityPrivacy
begin

builtins: hashing

/* 1. Setup: Generate the private identities */
rule Setup:
    [ Fr(~U), Fr(~V) ]
  --[ OnlyOnce() ]->
    [ !ID_U(~U), !ID_V(~V) ]

/* 2. Session One: Always uses Identity U */
rule Session_One:
    [ !ID_U(~U), Fr(~r1) ]
  -->
    [ Out(h(<~U, ~r1>)) ]

/* 3. Session Two: The 'Choice' rule
   Uses diff(U, V) to check for observational equivalence.
*/
rule Session_Two:
    [ !ID_U(~U), !ID_V(~V), Fr(~r2) ]
  -->
    [ Out(h(<diff(~U, ~V), ~r2>)) ]

/* --- Restrictions --- */
restriction OnlyOnce:
    "All #i #j. OnlyOnce() @ i & OnlyOnce() @ j ==> #i = #j"

end
\end{lstlisting}

\begin{figure}
    \centering
    \includegraphics[width=\linewidth]{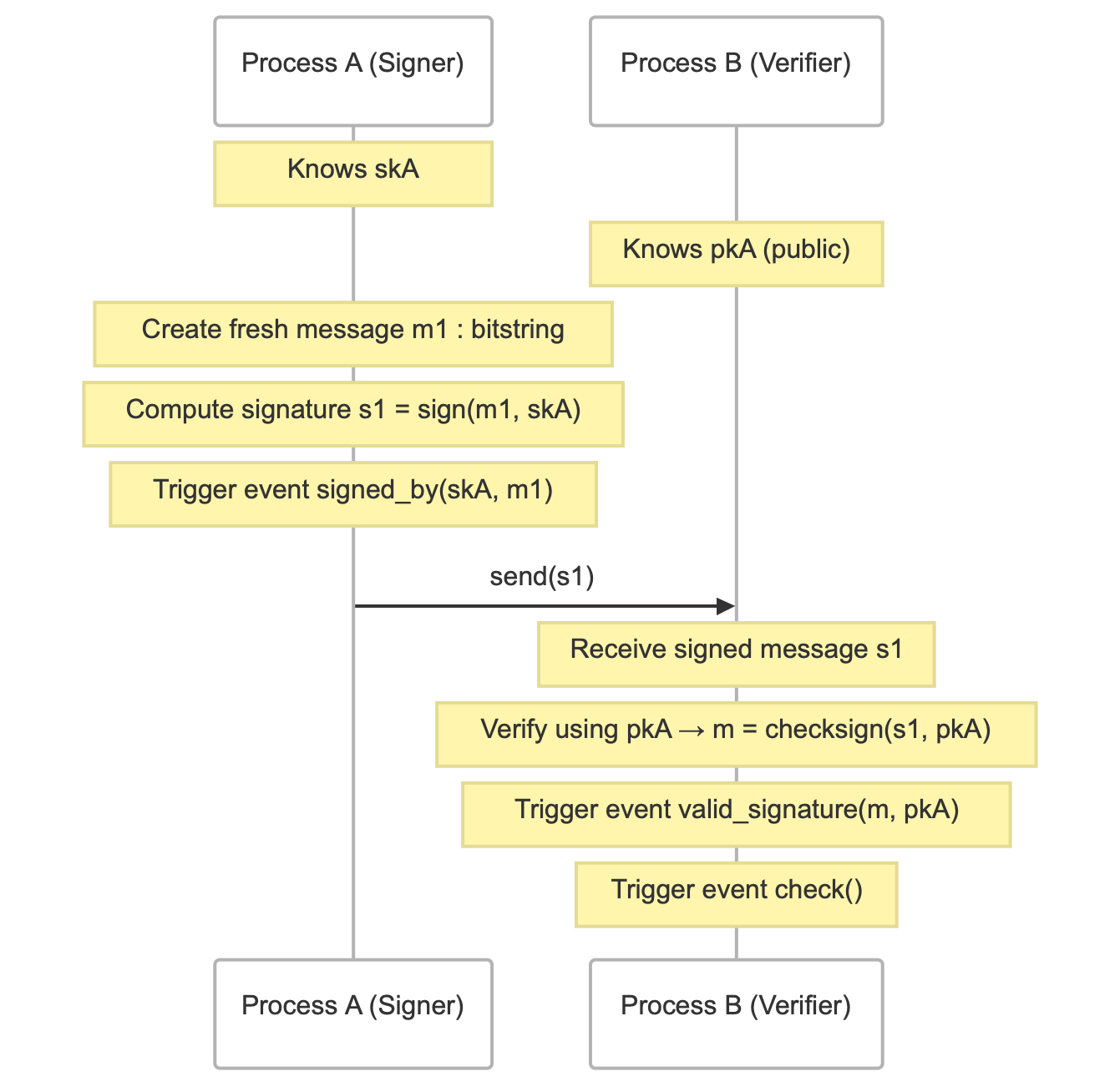}
    \caption{Example Protocol for Unforgeability Property}
    \label{fig:diagram-unforgeability}
\end{figure}

Unforgeability property ensures that no adversary can produce a valid message or signature that was not generated by an honest participant (see Section \ref{sec:results-rq2}). Figure \ref{fig:diagram-unforgeability} shows a sequence diagram of a simple protocol that achieves unforgeability~\cite{Lafourcade_Mahmoud_Marcadet_Olivier-Anclin_2024}. In this protocol, a \textit{Signer} signs a message using its secret key, sends the signed message to the \textit{Verifier}. Then, \textit{Verifier} verifies the signature with the public key of \textit{Signer}. Listing \ref{listing:pv-unforgeability} presents the ProVerif model of this protocol, where we define functions for the signing and verification process of the signature. We use these functions to model the protocol described by the sequence diagram and introduce some events to allow us to verify the correctness of the protocol. The ProVerif model triggers a \textit{signedBy} event with the secret key and the message after the \textit{Signer} signs the message, then it creates an event after the verification of the signature with the message and the public key of the signer as parameters. These two events enable us to define a query verifying that for each valid signature with a public key, there has been created a signature with corresponding secret key for that message. Conversely, Listing \ref{listing:tamarin-unforgeability} presents the Tamarin model of the same protocol, where we defined a lemma for the unforgeability check based on the formal definition from our taxonomy. The lemma states that for each valid signature, there is an event where an honest participant signed the message. In a similar manner, we have introduced \textit{SignedBy} and \textit{ValidSignature} events to allow us to apply the unforgeability lemma. 

\begin{lstlisting}[language=proverif, caption=Unforgeability Check in ProVerif, label=listing:pv-unforgeability]
free c: channel.

type skey.
type pkey.
type boolean.

fun pk(skey): pkey.
fun sign(bitstring, skey): bitstring.
reduc forall m: bitstring, sk: skey; checksign(sign(m, sk), pk(sk)) = m.

event signed_by(skey, bitstring).
event valid_signature(bitstring, pkey).
event check().

query m: bitstring, sk: skey; event(valid_signature(m, pk(sk))) ==> event(signed_by(sk, m)).
query event(check()).

free skA: skey [private].

let processA() =
  new m1: bitstring;
  let s1 = sign(m1, skA) in
  event signed_by(skA, m1);
  out(c, m1);
  out(c, s1).

let processB(pkA: pkey) =
  new m2: bitstring;
  in(c, signed_message: bitstring);
  let m = checksign(signed_message, pkA) in

  event valid_signature(m, pkA);
  event check();
  0.

process
  let pkA = pk(skA) in
  out(c, pkA);
  (!processA() | !processB(pkA))
\end{lstlisting}

\begin{lstlisting}[language=tamarin, caption=Unforgeability Check in Tamarin, label=listing:tamarin-unforgeability]
theory SimpleSignature
begin

builtins: signing

/* 1. Setup: Generate keys and publish the public key */
rule Setup:
    [ Fr(~skA) ]
  --[ OnlyOnce() ]->
    [ !LtkA(~skA), !PkA(pk(~skA)), Out(pk(~skA)) ]

/* 2. Process A: Sign a fresh message */
rule ProcessA:
    [ !LtkA(~skA), Fr(~m1) ]
  --[ SignedBy(~skA, ~m1) ]->
    [ Out(<~m1, sign(~m1, ~skA)>) ]

/* 3. Process B: Receive message and signature, then verify */
rule ProcessB:
    [ In(<m, sig>)  
    , !PkA(pkA) 
    ]
  --[ Eq(verify(sig, m, pkA), true)
    , ValidSignature(m, pkA)
    , Check()
    ]->
    [ ]

/* --- Restrictions --- */

restriction Equality:
    "All x y #i. Eq(x, y) @ i ==> x = y"

restriction OnlyOnce:
    "All #i #j. OnlyOnce() @ i & OnlyOnce() @ j ==> #i = #j"

/* --- Lemmas --- */

lemma Unforgeability:
    all-traces
    "All m pkA #i. ValidSignature(m, pkA) @ i ==> 
        (Ex sk #j. SignedBy(sk, m) @ j & pk(sk) = pkA)"

lemma Executability:
    exists-trace
    "Ex #i. Check() @ i"

end
\end{lstlisting}

Together, the sequence diagrams, ProVerif models, and Tamarin models collected in the accompanying repository create a reusable catalogue of modeling patterns that serves as a blueprint for specifying and verifying security properties. The applicability of this catalogue to concrete security protocols is demonstrated in Section \ref{sec:case-study}.

\section{Case Study}
\label{sec:case-study}
To validate the applicability of the taxonomy and modeling patterns to cryptographic primitives that combine multiple security goals, we analyze the Signcryption scheme proposed by Zheng~\cite{Zheng_1998}. Signcryption is a cryptographic primitive that performs signature and encryption in a single logical step. This is often regarded as secure message transmission. However, using our taxonomy, we decompose it into two security properties that require distinct verification strategies. One property is secrecy, which is a well-established property within the automated verification domain. In this case, the message \textit{m} must remain secret from the attacker. The other property is unforgeability, which is part of the integrity in our taxonomy. This ensures that a message originates from the claimed sender. In the case of ProVerif, this is verified by using event correspondence. 

We developed two models to demonstrate the necessity of the taxonomy's modeling patterns (see \textit{Example} folder in our repository). The initial models implement only the algebraic logic of the protocol. They define the key derivation functions (\textit{kdf1, kdf2}) and the core algebraic reduction where the Receiver's calculation matches the Sender's (for the signcryption). This model includes a \textit{success} event as a sanity check to ensure that the two actors can communicate. Nevertheless, this reachability check does not prove security, it only proves that the protocol can reach the end.

To verify unforgeability and secrecy (as claimed by the signcryption protocol), we applied the pattern for unforgeability and secrecy from our repository to the initial model. Based on this example, we extract the query and the events used and identify where the events are triggered (as presented in Listing \ref{listing:pv-unforgeability}, lines 23 and 32). In this case, the events are triggered before the signed message is sent and after the signed message is verified. Listing \ref{listing:pv-unforgeability-applied} presents the events and queries extracted from the unforgeability pattern and introduced in the signcryption protocol to verify the secrecy of the message and the unforgeability of the signature. The results of the verification indicate that the signcryption protocol ensures secrecy of the message and unforgeability of the signature as claimed by the protocol definition. 

\begin{lstlisting}[language=proverif, caption=Unforgeability ProVerif Events and Queries, label=listing:pv-unforgeability-applied]
(* --- 4. Events and Queries --- *)
event signed_by(skey, bitstring).
event valid_signature(bitstring, pkey). (* Message m accepted as coming from pkey *)

(* Sanity check *)
event success().
query event(success()).

(* Confidentiality: Can the attacker get m_0? *)
query attacker(m_0).

(* Unforgeability *)
query m: bitstring, sk: skey; event(valid_signature(m, pk(sk))) ==> event(signed_by(sk, m)).
\end{lstlisting}

In this case study, we followed the modeling pattern for unforgeability provided in the companion repository. The applicability of this pattern has been demonstrated on a security protocol by introducing the verification steps to both ProVerif and Tamarin models of the protocol. Even though the unforgeability definitions are available already, missing modeling patterns for the verification significantly increase the difficulty of modeling and verifying such security protocols. Therefore, this case study confirms that the taxonomy, along with the modeling patterns, provides a necessary bridge between abstract cryptographic definitions and concrete ProVerif and Tamarin implementations, even for schemes involving more complex algebraic reductions compared with the initial signing function provided in the pattern.

\section{Discussion}\label{sec:discussion}

The results reveal that confidentiality, authentication, and privacy are the most frequently verified properties, reflecting the maturity of tool support for these primitives. In contrast, accountability properties are rarely modeled, suggesting either an open research gap in automating their verification or that current security protocols do not prioritize this.

Unlike prior conceptual classifications \cite{Focardi_Gorrieri_Martinelli_2004, Rouland_Hamid_Bodeveix_Filali_2019, Sayar_Messe_Ebersold_Bruel_2025, hermann_taxonomy_2025}, our taxonomy is based on empirical evidence from recent verification studies and bridges the gap between theoretical definitions and executable models. Although these classifications provide rigorous formalization, we focus on improving understandability and practical applicability. We achieve this by deriving our taxonomy from current verification trends and building a catalog to demonstrate how these properties can be used in practice. 

We provide definitions of the security properties and propose small protocols to showcase ProVerif and Tamarin models to verify all security properties. We first give a conceptual understanding of the properties, then we present concrete examples for each property by creating code examples. These code examples aim to help users encode the verification of these security properties in their protocols~\cite{10.1145/1062455.1062491}. For example, in Section \ref{sec:results-rq3} we introduce an example for unforgeability. Given that the user already has a protocol encoded in ProVerif (without verification queries), the user can employ the queries that we defined for unforgeability into their model. 

\subsection{Threats to Validity}
This section addresses threats to internal, external, and construct validity, as statistical conclusion validity is not applicable given the qualitative nature of our study.
To mitigate the threats to validity, we employed a systematic approach where we defined a protocol that specifies the search string, inclusion/exclusion criteria, and data extraction form. Because the inclusion and exclusion criteria were clearly defined and objective, selection bias was minimized. Some subjectivity may have been introduced when determining whether a study included a description of security properties and corresponding ProVerif or Tamarin models. However, this assessment was largely straightforward, as the presence or absence of these elements was usually explicit in the papers. Largely, because for any security protocol study, they include a description of the properties they aim for. 

This study focuses only on security properties verified using ProVerif or Tamarin, thereby excluding works employing other tools such as Scyther or AVISPA. Nevertheless, when defining and formalizing security properties for our taxonomy, we did not limit ourselves to the selected papers, but took into consideration various relevant studies from the literature that propose widely accepted definitions and formalizations. 

During normalization, terminology across papers (e.g., identity protection, strong user authentication) required manual consolidation, which may have introduced interpretation bias. Thus, subjective grouping decisions could have influenced the taxonomy's structure. 

Since our paper only covers the studies published from 2022 to 2025, earlier works that introduced foundational security properties or alternative modeling approaches might be missed. However, this limitation was mitigated by performing targeted searches for each identified security property to extract relevant definitions and formalizations from relevant sources.

\section{Conclusion}\label{sec:conclusion}

This study establishes an evidence-based taxonomy of security properties crucial for the formal verification of security protocols using automated tools, specifically ProVerif and Tamarin. By systematically analyzing recent literature, we categorized verified properties into five core groups: Authentication, Confidentiality, Integrity, Privacy, and Accountability. The quantitative analysis provided empirical insight into current verification trends, confirming that core properties such as Confidentiality and Privacy dominate academic attention. A notable finding is the pronounced research gap in formal verification techniques for Accountability properties, which requires modeling complex concepts like evidence generation and judgment rather than simple message flows.

Crucially, we provided unified informal and formal definitions (in first-order logic) for each property in the taxonomy and detailed executable modeling patterns and best practices for both tools. This combination of rigor and practicality distinguishes our work from earlier conceptual or functional classifications. The provision of an open repository containing these formal models addresses the difficulty faced by protocol designers, offering a concrete and repeatable starting point for their verification efforts.

The immediate outcome of this study is the definitive foundation required for the development of our planned Domain-Specific Language (DSL). Future work will focus on:
\begin{enumerate}
    \item \textbf{DSL Development:} Creating the DSL to automatically translate high-level protocol descriptions and security properties into executable models for both ProVerif and Tamarin, thereby abstracting the complexity of formal verification.
    \item \textbf{Taxonomy Expansion: }Expanding the taxonomy to include higher-level security mechanisms, such as key exchange and multifactor authentication, that are composed of the primitive properties identified in this study.
\end{enumerate}

\backmatter

\section*{Declarations}

This research has been funded by the European Union’s research and innovation
programme ENTRUST, under grant agreement No. 101095634.

\bibliography{sn-bibliography}

@article{blanchet2018proverif,
  title={ProVerif 2.00: automatic cryptographic protocol verifier, user manual and tutorial},
  author={Blanchet, Bruno and Smyth, Ben and Cheval, Vincent and Sylvestre, Marc},
  journal={Version from},
  volume={16},
  pages={05--16},
  year={2018}
}

@inproceedings{meier2013tamarin,
  title={The TAMARIN prover for the symbolic analysis of security protocols},
  author={Meier, Simon and Schmidt, Benedikt and Cremers, Cas and Basin, David},
  booktitle={Computer Aided Verification: 25th International Conference, CAV 2013, Saint Petersburg, Russia, July 13-19, 2013. Proceedings 25},
  pages={696--701},
  year={2013},
  organization={Springer}
}

@inproceedings{cremers2008scyther,
  title={The scyther tool: Verification, falsification, and analysis of security protocols: Tool paper},
  author={Cremers, Cas JF},
  booktitle={International conference on computer aided verification},
  pages={414--418},
  year={2008},
  organization={Springer}
}

@InProceedings{AVISPA,
author="Armando, A.
and Basin, D.
and Boichut, Y.
and Chevalier, Y.
and Compagna, L.
and Cuellar, J.
and Drielsma, P. Hankes
and He{\'a}m, P. C.
and Kouchnarenko, O.
and Mantovani, J.
and M{\"o}dersheim, S.
and von Oheimb, D.
and Rusinowitch, M.
and Santiago, J.
and Turuani, M.
and Vigan{\`o}, L.
and Vigneron, L.",
editor="Etessami, Kousha
and Rajamani, Sriram K.",
title="The AVISPA Tool for the Automated Validation of Internet Security Protocols and Applications",
booktitle="Computer Aided Verification",
year="2005",
publisher="Springer Berlin Heidelberg",
address="Berlin, Heidelberg",
pages="281--285",
isbn="978-3-540-31686-2"
}

@article{KITCHENHAM20097-guidelines-slr,
title = {Systematic literature reviews in software engineering – A systematic literature review},
journal = {Information and Software Technology},
volume = {51},
number = {1},
pages = {7-15},
year = {2009},
note = {Special Section - Most Cited Articles in 2002 and Regular Research Papers},
issn = {0950-5849},
doi = {https://doi.org/10.1016/j.infsof.2008.09.009},
url = {https://www.sciencedirect.com/science/article/pii/S0950584908001390},
author = {Barbara Kitchenham and O. {Pearl Brereton} and David Budgen and Mark Turner and John Bailey and Stephen Linkman}
}

@inproceedings{Lowe_1997, address={Rockport, MA, USA}, title={A hierarchy of authentication specifications}, ISBN={978-0-8186-7990-2}, url={http://ieeexplore.ieee.org/document/596782/}, DOI={10.1109/CSFW.1997.596782}, abstractNote={Many security protocols have the aim of authenticating one agent to another Yet there is no clear consensus in the academic literature about precisely what “authentication” means. In this paper we suggest that the appropriate authentication requirement will depend upon the use to which the protocol is put, and identify several possible dejnitions of “authentication”. Weformalize each definition using the process algebra CSe use this formalism to study their relative strengths, and show how the model checker FDR can be used to test whether a system running the protocol meets such a speciJication.}, booktitle={Proceedings 10th Computer Security Foundations Workshop}, publisher={IEEE Comput. Soc. Press}, author={Lowe, G.}, year={1997}, pages={31–43}, language={en} }

@article{Biba_1977, title={Integrity Considerations for Secure Computer Systems}, number={522}, author={Biba, K J}, year={1977}, language={en} }

@inproceedings{Ahn_Kwak_Kim_2024, title={mdTLS: How to Make Middlebox-Aware TLS More Efficient?}, volume={14562}, ISBN={0302-9743}, archiveLocation={WOS:001209294600003}, DOI={10.1007/978-981-97-1238-0_3}, author={Ahn, T and Kwak, J and Kim, S}, editor={Seo, H and Kim, S}, year={2024}, pages={39–59}, booktitle={Information Security and Cryptology – ICISC 2023} }

@InProceedings{10.1007/3-540-44706-7_20,
author="Katz, Jonathan
and Yung, Moti",
editor="Goos, Gerhard
and Hartmanis, Juris
and van Leeuwen, Jan
and Schneier, Bruce",
title="Unforgeable Encryption and Chosen Ciphertext Secure Modes of Operation",
booktitle="Fast Software Encryption",
year="2001",
publisher="Springer Berlin Heidelberg",
address="Berlin, Heidelberg",
pages="284--299",
isbn="978-3-540-44706-1"
}

@inproceedings{Clement_Junqueira_Kate_Rodrigues_2012, address={Madeira Portugal}, title={On the (limited) power of non-equivocation}, ISBN={978-1-4503-1450-3}, url={https://dl.acm.org/doi/10.1145/2332432.2332490}, DOI={10.1145/2332432.2332490}, abstractNote={In recent years, there have been a few proposals to add a small amount of trusted hardware at each replica in a Byzantine fault tolerant system to cut back replication factors. These trusted components eliminate the ability for a Byzantine node to perform equivocation, which intuitively means making conﬂicting statements to diﬀerent processes. In this paper, we deﬁne non-equivocation and study its power in the context of distributed protocols that assume a Byzantine fault model. We show that non-equivocation alone does not allow for reducing the number of processes required to reach agreement in the presence of Byzantine faults in the asynchronous communication model, by proving a lower bound of n > 3f processes for agreement with nonequivocation. However, when we add the ability to guarantee the transferable authentication of network messages (e.g., using digital signatures), we show that it is possible to use non-equivocation to transform any protocol that works under the crash fault model into a protocol that tolerates Byzantine faults, without requiring an increase in the number of processes.}, booktitle={Proceedings of the 2012 ACM symposium on Principles of distributed computing}, publisher={ACM}, author={Clement, Allen and Junqueira, Flavio and Kate, Aniket and Rodrigues, Rodrigo}, year={2012}, month=july, pages={301–308}, language={en} }

@article{Abadi, title={Analyzing security protocols with secrecy types and logic programs}, abstractNote={We study and further develop two language-based techniques for analyzing security protocols. One is based on a typed process calculus; the other, on untyped logic programs. Both focus on secrecy properties. We contribute to these two techniques, in particular by extending the former with a ﬂexible, generic treatment of many cryptographic operations. We also establish an equivalence between the two techniques.}, author={Abadi, Martin and Blanchet, Bruno}, language={en} }

@inbook{Krawczyk_2021, address={Berlin, Heidelberg}, title={Perfect Forward Secrecy}, ISBN={978-3-642-27739-9}, DOI={10.1007/978-3-642-27739-9_90-2}, booktitle={Encyclopedia of Cryptography, Security and Privacy}, publisher={Springer Berlin Heidelberg}, author={Krawczyk, Hugo}, editor={Jajodia, Sushil and Samarati, Pierangela and Yung, Moti}, year={2021}, pages={1–3}, language={en} }

@article{Cremers_Jacomme_Naska, title={CISPA Helmholtz Center for Information Security}, abstractNote={The building blocks for secure messaging apps, such as Signal’s X3DH and Double Ratchet (DR) protocols, have received a lot of attention from the research community. They have notably been proved to meet strong security properties even in the case of compromise such as Forward Secrecy (FS) and Post-Compromise Security (PCS). However, there is a lack of formal study of these properties at the application level. Whereas the research works have studied such properties in the context of a single ratcheting chain, a conversation between two persons in a messaging application can in fact be the result of merging multiple ratcheting chains.}, author={Cremers, Cas and Jacomme, Charlie and Naska, Aurora}, language={en} }

@inproceedings{Cremers_Jacomme_Naska_2023, title={Formal Analysis of Session-Handling in Secure Messaging: Lifting Security from Sessions to Conversations}, volume={2}, author={Cremers, C. and Jacomme, C. and Naska, A.}, year={2023}, pages={1235–1252} }

@inbook{Schneider_Sidiropoulos_1996, address={Berlin, Heidelberg}, series={Lecture Notes in Computer Science}, title={CSP and anonymity}, volume={1146}, ISBN={978-3-540-61770-9}, DOI={10.1007/3-540-61770-1_38}, abstractNote={Security protocols are designed to meet particular security properties. In order to analyse such protocols formally, it is necessary to provide a formal definition of the property that they are intended to provide. This paper is concerned with the property of anonymity, tt proposes a definition of anonymity within the CSP notation, discusses the approach taken by CSP to anonymity with respect to different viewpoints, and illustrates this approach on some toy examples, and then applies it to a machine-assisted analysis of the dining cryptographers example and some variants.}, booktitle={Computer Security — ESORICS 96}, publisher={Springer Berlin Heidelberg}, author={Schneider, Steve and Sidiropoulos, Abraham}, editor={Bertino, Elisa and Kurth, Helmut and Martella, Giancarlo and Montolivo, Emilio}, year={1996}, pages={198–218}, collection={Lecture Notes in Computer Science}, language={en} }

@inproceedings{Samfat_Molva_Asokan_1995, address={Berkeley, California, United States}, title={Untraceability in mobile networks}, rights={https://www.acm.org/publications/policies/copyright_policy#Background}, ISBN={978-0-89791-814-5}, url={http://portal.acm.org/citation.cfm?doid=215530.215548}, DOI={10.1145/215530.215548}, abstractNote={User mobility is a feature that raises many new security-related issues and concerns. One of them is the disclosure of a mobile user’s reai identity during the authentication process, or other procedures specific to mobile networks. Such disclosure allows an unauthorized third-party to track the m.obile user’s movements and current whereabouts. Depending ou the context, access to auy information related to a mobile user’s location without his consent can be a serious violation of his privacy. This new issue might be seen as a conflicting requirement with respect to authentication: untraceability requires hiding the user’s identity while authentication requires the user’s identity to be revealed in order to be proved. What is needed is a single mechanism reconciling both authentication and privacy of a mobile user’s identification. The basic :solution to this problem is the use of uliases. Aliases insure untraceability by hiding the user’s real identity as well as his relationship with domain authorities. In this paper, we present a classification scheme to identify the various degrees of untraceability requirements. We then present an efficient method for the computation of aliases and apply It to a new set of inter-domain authentication protocols. We demonstrate that these protocols can be designed to meet various degrees of untraceability requirements. In designing these protocols, we try to avoid the drawbacks of authentication protoc:ols in existing mobile network architectures such as CDPD and GSM.}, booktitle={Proceedings of the 1st annual international conference on Mobile computing and networking  - MobiCom ’95}, publisher={ACM Press}, author={Samfat, Didier and Molva, Refik and Asokan, N.}, year={1995}, pages={26–36}, language={en} }

@inproceedings{Baelde_Delaune_Moreau_2020, title={A Method for Proving Unlinkability of Stateful Protocols}, volume={2020-June}, DOI={10.1109/CSF49147.2020.00020}, author={Baelde, D. and Delaune, S. and Moreau, S.}, year={2020}, pages={169–183} }

@article{Fu_Qin_Wang_Li_Zhang_2017, title={Nframe: A privacy-preserving with non-frameability handover authentication protocol based on (t, n) secret sharing for LTE/LTE-A networks}, volume={23}, ISSN={1022-0038, 1572-8196}, DOI={10.1007/s11276-016-1277-0}, abstractNote={Seamless handover between the evolved universal terrestrial radio access network and other access networks is highly desirable to mobile equipments in the long term evolution (LTE) or LTE-Advanced (LTE-A) networks, but ensuring security and efﬁciency of this process is challenging. In this paper, we propose a novel privacy-preserving with non-frameability handover authentication protocol based on (t, n) secret sharing to ﬁt in with all of the mobility scenarios in the LTE/LTE-A networks, which is called Nframe. To the best of our knowledge, Nframe is the ﬁrst to support protecting users’ privacy with non-frameability in the handover process. Moreover, Nframe uses pairing-free identity based cryptographic method to secure handover process and to achieve high efﬁciency. The formal veriﬁcation by the AVISPA tool shows that Nframe is secure against various malicious attacks and the simulation result indicates that it outperforms the existing schemes in terms of computation and communication cost.}, number={7}, journal={Wireless Networks}, author={Fu, Anmin and Qin, Ningyuan and Wang, Yongli and Li, Qianmu and Zhang, Gongxuan}, year={2017}, month=oct, pages={2165–2176}, language={en} }

@article{Xie_Wu_Hou_2023, title={BEPHAP: A Blockchain-based Efficient Privacy-Preserving Handover Authentication Protocol with key agreement for Internet of Vehicles}, volume={138}, url={https://www.scopus.com/inward/record.uri?eid=2-s2.0-85151687456&doi=10.1016%2fj.sysarc.2023.102869&partnerID=40&md5=e7805c8b60a279221dbe1e3279cbbff2}, DOI={10.1016/j.sysarc.2023.102869}, journal={Journal of Systems Architecture}, author={Xie, X. and Wu, B. and Hou, B.}, year={2023} }

@inproceedings{Schneider_1998, address={Rockport, MA, USA}, title={Formal analysis of a non-repudiation protocol}, ISBN={978-0-8186-8488-3}, url={http://ieeexplore.ieee.org/document/683155/}, DOI={10.1109/CSFW.1998.683155}, abstractNote={This paper applies the theory of Communicating Sequential Processes  CSP  to the modelling and analysis of a non-repudiation protocol. Non-repudiation protocols di er from authentication and key-exchange protocols in that the participants require protection from each other, rather than from an external hostile agent. This means that the kinds of properties that are required of such a protocol, and the way it needs to be modelled to enable analysis, are di erent to the standard approaches taken to the more widely studied class of protocols and properties. A non-repudiation protocol proposed by Zhou and Gollmann is analysed within this framework, and this highlights some novel considerations that are required for this kind of protocol.}, booktitle={Proceedings. 11th IEEE Computer Security Foundations Workshop (Cat. No.98TB100238)}, publisher={IEEE Comput. Soc}, author={Schneider, S.}, year={1998}, pages={54–65}, language={en} }

@book{Westin2015PrivacyAndFreedom,
  author    = {Westin, Alan F.},
  title     = {Privacy and Freedom},
  publisher = {Ig Publishing},
  address   = {Brooklyn, NY},
  year      = {2015},
  isbn      = {978-1-63246-073-8},
  note      = {With a new introduction by Daniel J. Solove}
}

@article{Ahmed_Peltonen_Sethi_Aura_2024, title={Security Analysis of the Consumer Remote SIM Provisioning Protocol}, volume={27}, url={https://doi.org/10.1145/3663761}, DOI={10.1145/3663761}, abstractNote={Remote SIM provisioning (RSP) for consumer devices is the protocol specified by the GSM Association for downloading SIM profiles into a secure element in a mobile device. The process is commonly known as eSIM, and it is expected to replace removable SIM cards. The security of the protocol is critical because the profile includes the credentials with which the mobile device will authenticate to the mobile network. In this article, we present a formal security analysis of the consumer RSP protocol. We model the multi-party protocol in applied pi calculus, define formal security goals, and verify them in ProVerif. The analysis shows that the consumer RSP protocol protects against a network adversary when all the intended participants are honest. However, we also model the protocol in realistic partial compromise scenarios where the adversary controls a legitimate participant or communication channel. The security failures in the partial compromise scenarios reveal weaknesses in the protocol design. The most important observation is that the security of RSP depends unnecessarily on it being encapsulated in a TLS tunnel. Also, the lack of pre-established identifiers means that a compromised download server anywhere in the world or a compromised secure element can be used for attacks against RSP between honest participants. Additionally, the lack of reliable methods for verifying user intent can lead to serious security failures. Based on the findings, we recommend practical improvements to RSP implementations, future versions of the specification, and mobile operator processes to increase the robustness of eSIM security.}, number={3}, journal={ACM Trans. Priv. Secur.}, publisher={Association for Computing Machinery}, author={Ahmed, Abu Shohel and Peltonen, Aleksi and Sethi, Mohit and Aura, Tuomas}, year={2024} }

@article{Akman_Ginzboorg_Damir_Niemi_2023, title={Privacy-Enhanced AKMA for Multi-Access Edge Computing Mobility †}, volume={12}, url={https://www.scopus.com/inward/record.uri?eid=2-s2.0-85146655686&doi=10.3390%2fcomputers12010002&partnerID=40&md5=f55536c91830a8c5ca15882be4fdf35b}, DOI={10.3390/computers12010002}, number={1}, journal={Computers}, author={Akman, G. and Ginzboorg, P. and Damir, M.T. and Niemi, V.}, year={2023} }

@inproceedings{B.Fila_S.Radomirović_2024, title={Nothing is Out-of-Band: Formal Modeling of Ceremonies}, ISBN={2374-8303}, DOI={10.1109/CSF61375.2024.00049}, note={journalAbbreviation: 2024 IEEE 37th Computer Security Foundations Symposium (CSF)}, booktitle={2024 IEEE 37th Computer Security Foundations Symposium (CSF)}, author={B. Fila and S. Radomirović}, year={2024}, month=july, pages={464–478} }

@inproceedings{Baloglu_Bursuc_Mauw_Pang_2024, series={ASIA CCS ’24}, title={Formal Verification and Solutions for Estonian E-Voting}, ISBN={979-8-4007-0482-6}, url={https://doi.org/10.1145/3634737.3657009}, DOI={10.1145/3634737.3657009}, abstractNote={Estonia has been deploying electronic voting for its government elections since 2005. The underlying e-voting system and protocol have been continuously improved, aiming to fix the vulnerabilities found over the years and to provide election verifiability, which is now the standard way to ensure election integrity despite corrupt infrastructure or parties. Another goal is receipt-freeness, to ensure privacy even if voters are coerced. However, several recent attacks against its verifiability and privacy show the need of rigorous, realistic formal specifications for the protocol and its security, of new solutions to mitigate attacks, and of automated security proofs to ensure all attacks have been covered. In this paper we propose:• a formal specification of the Estonian E-Voting protocol in a symbolic model suited for automated verification tools;• a general symbolic model for specifying privacy and receipt-freeness in presence of corrupt parties and infrastructure;• automated verification of security with respect to an exhaustive set of corruption scenarios, discovering new attacks on verifiability (with Tamarin) and on privacy (with ProVerif).• new solutions, focused on practical deployment and ease of use, and their automated proofs of security.}, publisher={Association for Computing Machinery}, author={Baloglu, Sevdenur and Bursuc, Sergiu and Mauw, Sjouke and Pang, Jun}, year={2024}, pages={728–741}, collection={ASIA CCS ’24} }

@article{Bodei_DeVincenzi_Matteucci_2024, title={Formal analysis of an AUTOSAR-based basic software module}, volume={26}, DOI={10.1007/s10009-024-00759-w}, number={4}, journal={International Journal on Software Tools for Technology Transfer}, author={Bodei, C. and De Vincenzi, M. and Matteucci, I.}, year={2024}, pages={495–508} }

@inproceedings{Bursuc_Horne_Mauw_Yurkov_2023, series={CCS ’23}, title={Provably Unlinkable Smart Card-based Payments}, ISBN={979-8-4007-0050-7}, url={https://doi.org/10.1145/3576915.3623109}, address = {New York, NY, USA}, DOI={10.1145/3576915.3623109}, abstractNote={The most prevalent smart card-based payment method, EMV, currently offers no privacy to its users. Transaction details and the card number are sent in cleartext, enabling the profiling and tracking of cardholders. Since public awareness of privacy issues is growing and legislation, such as GDPR, is emerging, we believe it is necessary to investigate the possibility of making payments anonymous and unlikable without compromising essential security guarantees and functional properties of EMV. This paper draws attention to trade-offs between functional and privacy requirements in the design of such a protocol. We present the UTX protocol - an enhanced payment protocol satisfying such requirements, and we formally certify key security and privacy properties using techniques based on the applied π-calculus.}, publisher={Association for Computing Machinery}, author={Bursuc, Sergiu and Horne, Ross and Mauw, Sjouke and Yurkov, Semen}, year={2023}, pages={1392–1406}, collection={CCS ’23} }

@article{Chunka_Banerjee_SachinKumar_2023, title={A secure communication using multifactor authentication and key agreement techniques in internet of medical things for COVID-19 patients}, volume={35}, url={https://www.scopus.com/inward/record.uri?eid=2-s2.0-85145566160&doi=10.1002%2fcpe.7602&partnerID=40&md5=253fa7789f795910b0345e0b45bd16b9}, DOI={10.1002/cpe.7602}, number={7}, journal={Concurrency and Computation: Practice and Experience}, author={Chunka, C. and Banerjee, S. and Sachin Kumar, G.}, year={2023} }

@inproceedings{DeVaere_Stoger_Perrig_Tsudik_2024, series={ASIA CCS ’24}, title={The SA4P Framework: Sensing and Actuation as a Privilege}, ISBN={979-8-4007-0482-6}, url={https://doi.org/10.1145/3634737.3657006}, address = {New York, NY, USA},DOI={10.1145/3634737.3657006}, abstractNote={Popular consumer Internet of Things (IoT) devices provide increasingly diverse sensing and actuation capabilities. Despite their benefits, such devices prompt numerous security concerns. Typically, security is attained at device-level granularity, which relies upon device trustworthiness. However, if a device is compromised (e.g., via remote exploits), this approach fails. To this end, we construct SA4P: Sensing and Actuation as a Privilege, a framework that decouples IoT devices from their physical environment. In SA4P, whenever any software on a device wants to access a sensing or actuation peripheral, it must be authorized to do so. This is achieved by the inclusion of an on-board component, Peripheral Guard (PEG), that physically guards peripherals. Besides providing strong security guarantees, SA4P motivates developers to consider sensing and actuation as valuable resources. SA4P’ design is modular, lightweight, and formally verified. It also does not require any hardware modifications for trusted execution environment (TEE)-equipped devices, while imposing only modest changes for other devices.}, publisher={Association for Computing Machinery}, author={De Vaere, Piet and Stoger, Felix and Perrig, Adrian and Tsudik, Gene}, year={2024}, pages={873–885}, collection={ASIA CCS ’24} }

@inproceedings{Duttagupta_Marin_Singelee_Preneel_2023, series={CODASPY ’23}, title={HAT: Secure and Practical Key Establishment for Implantable Medical Devices}, ISBN={979-8-4007-0067-5}, url={https://doi.org/10.1145/3577923.3583646}, address = {New York, NY, USA}, DOI={10.1145/3577923.3583646}, abstractNote={During the last few years, Implantable Medical Devices (IMDs) have evolved considerably. IMD manufacturers are now starting to rely on standard wireless technologies for connectivity. Moreover, there is an evolution towards open systems where the IMD can be remotely monitored or reconfigured through personal commercial-off-the-shelf devices such as smartphones or tablets. Nevertheless, a major problem that still remains unsolved today is the secure establishment of cryptographic keys between the IMD and such personal devices. Researchers have already proposed various solutions, most notably by relying on an additional external device. Unfortunately, these proposed approaches are either insecure, difficult to realise in practice, or are unsuitable for the latest generation of IMDs. Motivated by this, we present HAT, a secure and practical solution to provide fine-grained and dynamic access control for the next generation of IMDs, while offering full control and transparency to the patient. The main idea behind HAT is to shift the access control responsibilities from the IMD to an external device under the user’s control, such as a smartphone, acting as the IMD’s Key Distribution Center. We show that HAT only introduces minimal energy and memory overhead and formally prove its security using Verifpal.}, publisher={Association for Computing Machinery}, author={Duttagupta, Sayon and Marin, Eduard and Singelee, Dave and Preneel, Bart}, year={2023}, pages={213–224}, collection={CODASPY ’23} }

@inproceedings{Feng_Wang_Bobda_2025, series={FPGA ’25}, title={CIVIC-FPGA: A Trusted FPGA Design Validation by Multi-Tenant Cloud Providers}, ISBN={979-8-4007-1396-5}, url={https://doi.org/10.1145/3706628.3708826}, DOI={10.1145/3706628.3708826}, abstractNote={FPGA-based cloud services face significant security challenges, especially in multi-tenant environments where protecting tenant Intellectual Property (IP) is crucial. Existing solutions have limitations, being vulnerable to attacks or requiring FPGA manufacturer involvement for each deployment. We propose CIVIC-FPGA (Confidential IP Validation In Cloud for FPGA), a secure protocol for validating FPGA designs in multi-tenant cloud environments, ensuring tenant design confidentiality. CIVIC-FPGA validates tenant designs within cloud-provided Trusted Execution Environments (TEEs), using Public Key Infrastructure (PKI) to authenticate entities and verify TEE integrity. It creates a trusted link from the FPGA to its manufacturer, using device-specific keys and key exchanges to ensure that designs remain confidential and secure during validation and deployment. The protocol enables mutual authentication between tenants and FPGAs, supports encrypted bitstream transmission, and eliminates the need for FPGA manufacturer involvement in each deployment, thereby enhancing scalability. Furthermore, formal verification using ProVerif has been applied to rigorously validate the security properties of the protocol, confirming its resilience against key threats such as design tampering and replay attacks.  Benchmark tests show CIVIC-FPGA incurs minimal performance overhead, with 6% for simple designs and up to 17.5% for complex ones, providing a scalable and secure solution for multi-tenant cloud FPGA environments with minimal trade-offs.}, publisher={Association for Computing Machinery}, author={Feng, Yu and Wang, Zhaoqi and Bobda, Christophe}, year={2025}, pages={139–145}, collection={FPGA ’25} }

@inproceedings{Fujita_Yoneyama_2024, series={APKC ’24}, title={Formal Verification of Wireless Charging Standard Qi}, url={https://doi.org/10.1145/3659467.3659904}, address = {New York, NY, USA}, DOI={10.1145/3659467.3659904}, abstractNote={Wireless charging standard Qi is widely used, especially for smartphones, thanks to the convenience of charging the device simply by placing it on the charger. However, wireless charging has several security concerns. There is a great risk of making unintended connections because devices do not require any cable for connection. The privacy risk is also a concern because the charging history of devices such as smartphones and smartwatches may leak information like behavioral patterns if a charger can trace it. For these reasons, it is desirable to verify the security of wireless charging protocols formally. In this paper, we study the security of Qi with formal methods. For the verification, we use ProVerif, an automated cryptographic protocol verification tool. We verify secrecy of private keys, authenticity, replay attack resistance, and privacy in the taxonomic combination (507 patterns) of all multiple options of Qi execution. As a result of the verification, we find that the secrecy is satisfied in all situations, and there are only trivial attacks on authenticity and privacy.}, publisher={Association for Computing Machinery}, author={Fujita, Kazuhiro and Yoneyama, Kazuki}, year={2024}, pages={23–31}, collection={APKC ’24} }

@inproceedings{Giantsidi_Pritzi_Gust_Katsarakis_Koshiba_Bhatotia_2025, series={ASPLOS ’25}, title={TNIC: A Trusted NIC Architecture: A hardware-network substrate for building high-performance trustworthy distributed systems}, ISBN={979-8-4007-1079-7}, url={https://doi.org/10.1145/3676641.3716277}, DOI={10.1145/3676641.3716277}, abstractNote={We introduce TNIC, a trusted NIC architecture for building trustworthy distributed systems deployed in heterogeneous, untrusted (Byzantine) cloud environments. TNIC builds a minimal, formally verified, silicon root-of-trust at the network interface level. We strive for three primary design goals: (1) a host CPU-agnostic unified security architecture by providing trustworthy network-level isolation; (2) a minimalistic and verifiable TCB based on a silicon root-of-trust by providing two core properties of transferable authentication and non-equivocation; and (3) a hardware-accelerated trustworthy network stack leveraging SmartNICs. Based on the TNIC architecture and associated network stack, we present a generic set of programming APIs and a recipe for building high-performance, trustworthy, distributed systems for Byzantine settings. We formally verify the safety and security properties of our TNIC while demonstrating its use by building four trustworthy distributed systems. Our evaluation of TNIC shows up to 6texttimes performance improvement compared to CPU-centric TEE systems.}, publisher={Association for Computing Machinery}, author={Giantsidi, Dimitra and Pritzi, Julian and Gust, Felix and Katsarakis, Antonios and Koshiba, Atsushi and Bhatotia, Pramod}, year={2025}, pages={1282–1301}, collection={ASPLOS ’25} }

@article{H.Feng_J.Guan_H.Li_X.Pan_Z.Zhao_2023, title={FIDO Gets Verified: A Formal Analysis of the Universal Authentication Framework Protocol}, volume={20}, ISSN={1941-0018}, DOI={10.1109/TDSC.2022.3217259}, number={5}, journal={IEEE Transactions on Dependable and Secure Computing}, author={H. Feng and J. Guan and H. Li and X. Pan and Z. Zhao}, year={2023}, month=oct, pages={4291–4310} }

@inproceedings{Hu_Zhang_Weimerskirch_Mao_2022, series={ASIA CCS ’22}, title={Gatekeeper: A Gateway-based Broadcast Authentication Protocol for the In-Vehicle Ethernet}, ISBN={978-1-4503-9140-5}, url={https://doi.org/10.1145/3488932.3517396}, address = {New York, NY, USA}, DOI={10.1145/3488932.3517396}, abstractNote={Automotive Ethernet is considered to be the next-generation in-vehicle network, because of its high bandwidth, high throughput, and low cost characteristics. However, no common standard has been established for the security protocol of Automotive Ethernet. While there are a few candidates, including MACsec, IPsec, and TLS, there is no widely favored candidate. Most importantly, existing candidates cannot fully satisfy the requirements of in-vehicle communication, specifically source authentication for broadcast/multicast communication. In this paper, we conduct a comprehensive analysis in both security and performance of existing security protocol candidates and identify source authentication and Denial-of-Service (DoS) prevention as two essential but missing properties in these candidates. We propose Gatekeeper, a gateway-based broadcast authentication protocol to ensure source authentication. In general, Gatekeeper introduces an on-path authenticator, which co-locates with the in-vehicle gateway or domain controllers and helps receivers to verify the sender’s identity. To defend against DoS threats, we further integrate the time-lock puzzle with Gatekeeper to slow down malicious traffic. Our performance evaluation results show that Gatekeeper only results in 0.03 ms latency overhead for CAN data transmission and outperforms TESLA on both CAN and LiDAR transmission scenarios, highlighting the effectiveness and efficiency of Gatekeeper.}, publisher={Association for Computing Machinery}, author={Hu, Shengtuo and Zhang, Qingzhao and Weimerskirch, Andre and Mao, Z. Morley}, year={2022}, pages={494–507}, collection={ASIA CCS ’22} }

@inproceedings{Jacomme_Klein_Kremer_Racouchot_2023, title={A comprehensive, formal and automated analysis of the EDHOC protocol}, volume={8}, author={Jacomme, C. and Klein, E. and Kremer, S. and Racouchot, M.}, year={2023}, pages={5881–5898} }

@article{Kim_Ryu_Lee_Lee_Won_2023, title={Distributed and Federated Authentication Schemes Based on Updatable Smart Contracts}, volume={12}, url={https://www.scopus.com/inward/record.uri?eid=2-s2.0-85149642574&doi=10.3390%2felectronics12051217&partnerID=40&md5=2b5005860426e30d9b9d751717399150}, DOI={10.3390/electronics12051217}, number={5}, journal={Electronics (Switzerland)}, author={Kim, K. and Ryu, J. and Lee, H. and Lee, Y. and Won, D.}, year={2023} }

@article{Ko_Pawana_Won_Astillo_You_2024, title={Toward an Era of Secure 5G Convergence Applications: Formal Security Verification of 3GPP AKMA with TLS 1.3 PSK Option}, volume={14}, url={https://www.scopus.com/inward/record.uri?eid=2-s2.0-85211916802&doi=10.3390%2fapp142311152&partnerID=40&md5=58315b0d2281fc6d382744a0f7bef1cc}, DOI={10.3390/app142311152}, number={23}, journal={Applied Sciences (Switzerland)}, author={Ko, Y. and Pawana, I.W.A.J. and Won, T. and Astillo, P.V. and You, I.}, year={2024} }

@inproceedings{Lafourcade_Mahmoud_Marcadet_Olivier-Anclin_2024, series={ASIA CCS ’24}, title={Transferable, Auditable and Anonymous Ticketing Protocol}, ISBN={979-8-4007-0482-6}, url={https://doi.org/10.1145/3634737.3645008}, DOI={10.1145/3634737.3645008}, abstractNote={Digital ticketing systems typically offer ticket purchase, refund, validation, and, optionally, anonymity of users. However, it would be interesting for users to transfer their tickets, as is currently done with physical tickets. We propose Applause, a ticketing system allowing the purchase, refund, validation, and transfer of tickets based on trusted authority, while guaranteeing the anonymity of users, as long as the used payment method provides anonymity. To study its security, we formalise the security of the transferable E-Ticket scheme in the game-based paradigm. We prove the security of Applause computationally in the standard model and symbolically using the protocol verifier ProVerif. Applause relies on standard cryptographic primitives, rendering our construction efficient and scalable, as shown by a proof-of-concept. In order to obtain Spotlight, an auditable version, proved to be secure, users will remain anonymous except for a trusted third party, which will be able to disclose their identity in the event of a disaster.}, publisher={Association for Computing Machinery}, author={Lafourcade, Pascal and Mahmoud, Dhekra and Marcadet, Gael and Olivier-Anclin, Charles}, year={2024}, pages={1911–1927}, collection={ASIA CCS ’24} }

@article{Li_2023, title={A secure and efficient three-factor authentication protocol for IoT environments}, volume={179}, ISSN={0743-7315}, DOI={10.1016/j.jpdc.2023.104714}, abstractNote={The Internet of Things (IoT) is an information carrier based on the Internet and traditional telecommunications network, which enables all ordinary physical objects that can be independently addressed to form an interconnected network. User authentication protocol is an essential technology for security and privacy in the IoT environment. This paper analyzes the security of Mirsaraei et al.’s three-factor authentication scheme for IoT environments (Mirsaraei et al., 2022 [31]), and finds that the scheme cannot provide users with untraceability, perfect forward secrecy or the resistance of key compromise impersonation attack. The article improves Mirsaraei et al.’s scheme and proposes a three-factor authentication protocol with perfect forward secrecy using elliptic curve cryptosystem, which retains the general process of Mirsaraei et al.’s scheme. The formal security analysis of the proposed protocol is carried out by ROR (Real-or-Random) model, and the formal security verification of the proposed protocol is implemented by Proverif tool. The cryptoanalysis results demonstrate that the proposed protocol makes up for the shortcomings of Mirsaraei et al.’s scheme in security and can resist more malicious attacks as opposed to recent schemes. Moreover, the performance analysis using MIRACL (Multiprecision Integer and Rational Arithmetic C/C++ Library) shows that, the proposed protocol has great advantages over analogical three-factor authentication schemes in terms of computational overhead and communication overhead.}, journal={Journal of Parallel and Distributed Computing}, author={Li, Yi}, year={2023}, month=sept, pages={104714} }

@inproceedings{Li_Jin_Levchenko_2024, series={CCS ’24}, title={CAPSID: A Private Session ID System for Small UAVs}, ISBN={979-8-4007-0636-3}, url={https://doi.org/10.1145/3658644.3690324}, DOI={10.1145/3658644.3690324}, abstractNote={The US Federal Aviation Administration (FAA) has recently mandated that small unmanned aerial vehicles (UAVs) be equipped with a transmitter that broadcasts the UAV’s serial number, location, and altitude. The inclusion of a unique identifier in the form of a UAV serial number has stoked fears that the identifier will be used to track UAV operators and has even led some UAV operators to file a lawsuit against the FAA.In this paper, we propose CAPSID, an implementation of the FAA session ID concept that provides message authentication-something current Remote ID implementations lack-and strong operator anonymity. The FAA (or its equivalent in other jurisdictions) retains the ability to de-anonymize operators, but only in circumstances prescribed by law. To make this possible, CAPSID introduces a partially trusted third party, the Custodian, that serves as the bridge between anonymous identifiers and true operator identity. The Custodian ensures that the legal requirements for de-anonymization are met, preventing unnecessary mass collection of personally identifying information by a government agency.We formally verify the message authentication property of CAPSID using a cryptographic protocol checker and provide a formal proof of identifier non-linkability, even in the presence of corrupt (but non-colluding) authorities.}, publisher={Association for Computing Machinery}, author={Li, Yueshen and Jin, Jianli and Levchenko, Kirill}, year={2024}, pages={1791–1805}, collection={CCS ’24} }

@article{Liu_Zhou_Cao_Xu_Wang_Gao_Zeng_Xu_2023, title={A Robust and Effective Two-Factor Authentication (2FA) Protocol Based on ECC for Mobile Computing}, volume={13}, url={https://www.scopus.com/inward/record.uri?eid=2-s2.0-85152587051&doi=10.3390%2fapp13074425&partnerID=40&md5=61d64233d6c1aaa6121dabdea4b74bd6}, DOI={10.3390/app13074425}, number={7}, journal={Applied Sciences (Switzerland)}, author={Liu, K. and Zhou, Z. and Cao, Q. and Xu, G. and Wang, C. and Gao, Y. and Zeng, W. and Xu, G.}, year={2023} }

@article{M.Pradhan_S.Mohanty_2024, title={A Blockchain-Assisted Multifactor Authentication Protocol for Enhancing IoMT Security}, volume={11}, ISSN={2327-4662}, DOI={10.1109/JIOT.2024.3422242}, number={24}, journal={IEEE Internet of Things Journal}, author={M. Pradhan and S. Mohanty}, year={2024}, month=dec, pages={39323–39332} }

@article{Méré_Jouault_Pallardy_Perdriau_2024, title={Evaluating formal model verification tools in an industrial context: the case of a smart device life cycle management system}, url={https://www.scopus.com/inward/record.uri?eid=2-s2.0-85201418771&doi=10.1007%2fs10270-024-01201-0&partnerID=40&md5=8a8ac12fca33591416da8969578f24b7}, DOI={10.1007/s10270-024-01201-0}, journal={Software and Systems Modeling}, author={Méré, M. and Jouault, F. and Pallardy, L. and Perdriau, R.}, year={2024} }

@article{Miculan_Vitacolonna_2023, title={Automated verification of Telegram’s MTProto 2.0 in the symbolic model}, volume={126}, ISSN={0167-4048}, DOI={10.1016/j.cose.2022.103072}, abstractNote={MTProto 2.0 is the suite of security protocols for instant messaging at the core of the popular Telegram messenger application. In this paper we analyse MTProto 2.0 using ProVerif, a state-of-the-art symbolic security protocol verifier based on the Dolev–Yao model. We provide the first formal symbolic model of MTProto 2.0; in this model, we provide fully automated proofs of the soundness of authentication, normal chat, end-to-end encrypted chat, and rekeying mechanisms with respect to several security properties, including authentication, integrity, secrecy and perfect forward secrecy. At the same time, we discover that the rekeying protocol is vulnerable to an unknown key-share (UKS) attack. To achieve these results, we proceed in an incremental way: each protocol is examined in isolation, relying only on the guarantees provided by the previous ones and the robustness of the basic cryptographic primitives. The importance of this research is threefold. First, it proves the formal correctness of MTProto 2.0 with respect to most relevant security properties. Secondly, we isolate the aspects of cryptographic primitives that escape the symbolic model and thus require further investigation in the computational model. Finally, our modelisation can serve as a reference for the implementation and analysis of clients and servers.}, journal={Computers \& Security}, author={Miculan, Marino and Vitacolonna, Nicola}, year={2023}, month=mar, pages={103072} }

@inproceedings{Moustafa_Sethi_Aura_2025, title={Misbinding Raw Public Keys to Identities in TLS}, volume={15396 LNCS}, DOI={10.1007/978-3-031-79007-2\_4}, author={Moustafa, M. and Sethi, M. and Aura, T.}, year={2025}, pages={62–79} }

@inproceedings{Pietro_Savio_Roberto_2023, series={ACSAC ’23}, title={Lightweight Privacy-Preserving Proximity Discovery for Remotely-Controlled Drones}, ISBN={979-8-4007-0886-2}, url={https://doi.org/10.1145/3627106.3627174}, DOI={10.1145/3627106.3627174}, abstractNote={Discovering mutual proximity and avoiding collisions is one of the most critical services needed by the next generation of Unmanned Aerial Vehicles (UAVs). However, currently available solutions either rely on sharing mutual locations, neglecting the location privacy of involved parties, or are applicable for fully autonomous vehicles only—leaving unaddressed Remotely-Piloted UAVs’ safety needs. Alternatively, proximity can be discovered by adding sensing capabilities. However, in addition to the cost of the sensors, the complexity of integration, and the toll on the energy budget, the effectiveness of such solutions is usually limited by short detection ranges, making them hardly useful in high-mobility scenarios. In this paper, we propose LPPD (an acronym for Lightweight Privacy-preserving Proximity Discovery), a unique solution for privacy-preserving proximity discovery among remotely piloted UAVs based on the exchange of wireless messages. LPPD integrates two main building blocks: (i) a custom space tessellation technique based on randomized spheres; and, (ii) a lightweight cryptographic primitive for private-set intersection. Another feature enjoyed by LPPD is that it does not require online third parties. LPPD is rooted in sound theoretical results and is supported by an experimental assessment performed on a real drone. In particular, experimental results show that LPPD achieves 100% proximity discovery while taking only 39.66&nbsp;milliseconds in the most lightweight configuration and draining only the 5 · 10− 6% of the UAV’s battery capacity. In addition, LPPD’s security properties are formally verified.}, publisher={Association for Computing Machinery}, author={Pietro, Tedeschi and Savio, Sciancalepore and Roberto, Di Pietro}, year={2023}, pages={178–189}, collection={ACSAC ’23} }

@article{Raimondo_Bernardi_Marrone_Merseguer_2023, title={An approach for the automatic verification of blockchain protocols: the Tweetchain case study}, volume={19}, DOI={10.1007/s11416-022-00444-z}, number={1}, journal={Journal of Computer Virology and Hacking Techniques}, author={Raimondo, M. and Bernardi, S. and Marrone, S. and Merseguer, J.}, year={2023}, pages={17–32} }

@inproceedings{Rakeei_Giustolisi_Lenzini_2023, title={Secure Internet Exams Despite Coercion}, volume={13619 LNCS}, DOI={10.1007/978-3-031-25734-6_6}, author={Rakeei, M. and Giustolisi, R. and Lenzini, G.}, year={2023}, pages={85–100} }

@article{Rangwani_Om_2023, title={A Robust Four-Factor Authentication Protocol for Resource Mining}, volume={48}, DOI={10.1007/s13369-022-07055-2}, number={2}, journal={Arabian Journal for Science and Engineering}, author={Rangwani, D. and Om, H.}, year={2023}, pages={1947–1971} }

@inproceedings{S.Bussa_R.Sisto_F.Valenza_2023a, title={Formal Verification of a V2X Privacy Preserving Scheme Using Proverif}, DOI={10.1109/CSR57506.2023.10224908}, note={journalAbbreviation: 2023 IEEE International Conference on Cyber Security and Resilience (CSR)}, booktitle={2023 IEEE International Conference on Cyber Security and Resilience (CSR)}, author={S. Bussa and R. Sisto and F. Valenza}, year={2023}, month=aug, pages={341–346} }

@inproceedings{S.Bussa_R.Sisto_F.Valenza_2023b, title={Formal Verification of the FDO Protocol}, ISBN={2644-3252}, DOI={10.1109/CSCN60443.2023.10453172}, note={journalAbbreviation: 2023 IEEE Conference on Standards for Communications and Networking (CSCN)}, booktitle={2023 IEEE Conference on Standards for Communications and Networking (CSCN)}, author={S. Bussa and R. Sisto and F. Valenza}, year={2023}, month=nov, pages={290–295} }

@article{S.Lu_Z.Li_X.Miao_Q.Han_J.Zheng_2023, title={PIWS: Private Intersection Weighted Sum Protocol for Privacy-Preserving Score-Based Voting With Perfect Ballot Secrecy}, volume={10}, ISSN={2329-924X}, DOI={10.1109/TCSS.2022.3162869}, number={3}, journal={IEEE Transactions on Computational Social Systems}, author={S. Lu and Z. Li and X. Miao and Q. Han and J. Zheng}, year={2023}, month=june, pages={1039–1056} }

@inproceedings{Sabry_Samavi_2022, series={ACSAC ’22}, title={ArchiveSafe LT: Secure Long-term Archiving System}, ISBN={978-1-4503-9759-9}, url={https://doi.org/10.1145/3564625.3564635}, DOI={10.1145/3564625.3564635}, abstractNote={Every year the amount of digitally stored sensitive information increases significantly. Information such as governmental and legal documents, health, and tax records are required to be securely archived for decades to comply with various laws and regulations. Since cryptographic schemes based on single computational assumptions are not guaranteed to stay secure for such long periods, current state-of-the-art systems providing long-term confidentiality and integrity rely on information-theoretic techniques, such as multi-server secret sharing and commitments. These systems achieve the desired results; however, establishing private channels for secret sharing is costly and requires a complex setup. In this paper, we present ArchiveSafe LT, a framework for archiving systems aiming to provide long-term confidentiality and integrity. The framework relies on multiple computationally-secure schemes using robust combiners, with a design that plans for agility and evolution of cryptographic schemes. ArchiveSafe LT is efficient and suitable for practical adoption as it eliminates the need for private channels compared to its counterparts. We present the ArchiveSafe LT framework structure and its security analysis using an automatic prover. We specify two ArchiveSafe LT-based system designs, which handle different adversarial storage providers. We experimentally evaluate a prototype built based on one of the designs to show the system’s efficiency compared to information-theoretic systems.}, publisher={Association for Computing Machinery}, author={Sabry, Moe and Samavi, Reza}, year={2022}, pages={936–948}, collection={ACSAC ’22} }

@article{Saini_Kaur_Kumar_Kumar_2024, title={An efficient three-factor authentication protocol for wireless healthcare sensor networks}, volume={83}, DOI={10.1007/s11042-024-18114-1}, number={23}, journal={Multimedia Tools and Applications}, author={Saini, K.K. and Kaur, D. and Kumar, D. and Kumar, B.}, year={2024}, pages={63699–63721} }

@inproceedings{Seo_Kim_Lee_Kwon_Seo_2023, title={Efficient Remote Identification for Drone Swarms}, volume={76}, DOI={10.32604/cmc.2023.039459}, number={3}, author={Seo, K.-M. and Kim, J. and Lee, S. and Kwon, J.-W. and Seo, S.-H.}, year={2023}, pages={2937–2958} }

@article{Shahrouz_Analoui_2024, title={An anonymous authentication scheme with conditional privacy-preserving for Vehicular Ad hoc Networks based on zero-knowledge proof and Blockchain}, volume={154}, ISSN={1570-8705}, DOI={10.1016/j.adhoc.2023.103349}, abstractNote={Recently, Vehicular Ad hoc Networks (VANETs) have gained extensive attention in both academia and industry. VANETs play a seminal role in smart transportation by enhancing driving convenience and traffic efficiency through real-time information interaction. Despite this, ensuring secure authentication and privacy preservation remains two challenging issues in VANETs. The existence of security schemes that are vulnerable to privacy and security issues or have substantial computation and communication overheads shows the necessity for further research in this area. Therefore, in this article, we present an anonymous authentication scheme based on Zero-Knowledge Proof. To be more specific, we adopted Simulation Extractable Zero-Knowledge SNARKs (SE zk-SNARKs) to achieve anonymity and conditional privacy. Although zk-SNARK proofs are succinct and fast to verify, the time required for proof generation poses a major obstacle in using zk-SNARKs in a time-sensitive VANET environment. To overcome this drawback, we focus on key components of the proof statement, separating the repetitive calculations, and archiving in the Blockchain, which serves as an immutable data storage. By omitting repetitive calculations, our proposed scheme becomes lightweight enough to run efficiently on VANETs. Security analysis showed that the proposed protocol satisfied security and privacy requirements. Additionally, a computation cost analysis is provided to demonstrate the significant advantage of our proposed scheme. ProVerif is used to verify the strong secrecy of the protocol, and the results show that privacy can be guaranteed in the proposed scheme.}, journal={Ad Hoc Networks}, author={Shahrouz, Jamile Khalili and Analoui, Morteza}, year={2024}, month=mar, pages={103349} }

@inproceedings{Wagner_Birnstill_Beyerer_2024, series={ARES ’24}, title={DDS Security+: Enhancing the Data Distribution Service With TPM-based Remote Attestation}, ISBN={979-8-4007-1718-5}, url={https://doi.org/10.1145/3664476.3670442}, DOI={10.1145/3664476.3670442}, abstractNote={The Data Distribution Service (DDS) is a widely accepted industry standard for reliably exchanging data over the network using a publish-subscribe model. While DDS already includes basic security features such as participant authentication and access control, the possibilities of leveraging Trusted Platform Modules (TPMs) to increase the security and trustworthiness of DDS-based applications have not been sufficiently researched yet. In this work, we show how TPM-based remote attestation can be effectively integrated into the existing DDS security architecture. This enables application developers to verify the code integrity of remote DDS participants during the operation of the distributed system. Our solution transparently extends the DDS secure channel handshake, while cryptographically binding the established communication channels to the attested software stacks. We show the security properties of our proposal by formally verifying the resulting remote attestation protocol using the Tamarin theorem prover. We also implement our solution as a fork of the popular eProsima FastDDS library and evaluate the resulting performance impact when conducting TPM-based remote attestations of DDS applications.}, publisher={Association for Computing Machinery}, author={Wagner, Paul Georg and Birnstill, Pascal and Beyerer, Jurgen}, year={2024}, collection={ARES ’24} }

@inproceedings{Wang_Li_Guan_2024, series={ASIA CCS ’24}, title={A Formal Analysis of Data Distribution Service Security}, ISBN={979-8-4007-0482-6}, url={https://doi.org/10.1145/3634737.3656288}, address = {New York, NY, USA}, DOI={10.1145/3634737.3656288}, abstractNote={The Data Distribution Service (DDS) constructs a highly available data transmission middleware based on the publish-subscribe model, widely used in the Internet of Things environment. To improve the security of DDS, the Object Management Group formulated the DDS Security, which provides security mechanisms for DDS in the form of security plugins. However, the security of the DDS Security protocol has not been fully analyzed. We analyze DDS Security through formal methods. We model the security goals and protocol flow of the DDS Security using ProVerif and evaluate whether its security goals can be met in different scenarios. Our analysis confirms previously manually identified vulnerabilities in an automated way and reveals new attacks. We discovered the permission file impersonation attack, the denial of service attack, the degradation attack, and the privacy leakage attack guided by the formal analysis result. For these threats, we propose corresponding mitigation measures and recommendations.}, publisher={Association for Computing Machinery}, author={Wang, Binghan and Li, Hui and Guan, Jingjing}, year={2024}, pages={716–727}, collection={ASIA CCS ’24} }

@article{Wang_Wu_Wen_Zhou_Hu_Xie_2025, title={An Improved Blockchain-Based Lightweight Vehicle-to-Infrastructure Handover Authentication Protocol for Vehicular Ad Hoc Networks}, volume={13}, url={https://www.scopus.com/inward/record.uri?eid=2-s2.0-105002233804&doi=10.3390%2fmath13071118&partnerID=40&md5=16f366ddc3ea74b266ef1caf20a84627}, DOI={10.3390/math13071118}, number={7}, journal={Mathematics}, author={Wang, S. and Wu, Y. and Wen, K. and Zhou, X. and Hu, B. and Xie, Q.}, year={2025} }

@inproceedings{Wang_Laing_Moreira_Ryan_2024, series={CODASPY ’24}, title={Remote Registration of Multiple Authenticators}, ISBN={979-8-4007-0421-5}, url={https://doi.org/10.1145/3626232.3653273}, address = {New York, NY, USA}, DOI={10.1145/3626232.3653273}, abstractNote={User authentication with discrete authenticators, such as YubiKeys, is becoming increasingly popular. The authenticators can be external or on-device. They work using challenge-response protocols and public key cryptography. Multiple accounts can be associated with each authenticator. Compared with other forms of authentication, this approach has advantages in security and usability. There are, however, significant limitations which persist. In particular, if users possess only one authenticator, they lack resilience to loss and malfunction. On the other hand, if they possess multiple authenticators, they lack practical solutions to keep authenticators synchronised. In this paper, we present three solutions which combine the usability of a single authenticator with the resilience of multiple authenticators. We also present two key derivation functions which are used in our solutions. All three solutions maintain core security and privacy properties found in existing systems. Meanwhile, each solution provides additional value in different use cases. The security of our solutions is analysed using ProVerif.}, publisher={Association for Computing Machinery}, author={Wang, Yongqi and Laing, Thalia and Moreira, Jose and Ryan, Mark D.}, year={2024}, pages={379–390}, collection={CODASPY ’24} }

@inproceedings{Watanabe_Yoneyama_2024, title={Formal Verification of Challenge Flow in EMV 3-D Secure}, volume={14896 LNCS}, DOI={10.1007/978-981-97-5028-3_15}, author={Watanabe, K. and Yoneyama, K.}, year={2024}, pages={290–310} }

@article{Wu_Meng_Yang_Kumari_Pirouz_2022, title={Amassing the Security: An Enhanced Authentication and Key Agreement Protocol for Remote Surgery in Healthcare Environment}, volume={134}, ISSN={1526-1492}, DOI={10.32604/cmes.2022.019595}, abstractNote={The development of the Internet of Things has facilitated the rapid development of various industries. With the improvement in people’s living standards, people’s health requirements are steadily improving. However, owing to the scarcity of medical and health care resources in some areas, the demand for remote surgery has gradually increased. In this paper, we investigate remote surgery in the healthcare environment. Surgeons can operate robotic arms to perform remote surgery for patients, which substantially facilitates successful surgeries and saves lives. Recently, Kamil et al. proposed a secure protocol for surgery in the healthcare environment. However, after cryptanalyzing their protocol, we deduced that their protocols are vulnerable to temporary value disclosure and insider attacks. Therefore, we design an improved authentication and key agreement protocol for remote surgeries in the healthcare environment. Accordingly, we adopt the real or random (ROR) model and an automatic verification tool Proverif to verify the security of our protocol. Via security analysis and performance comparison, it is confirmed that our protocol is a relatively secure protocol.}, number={1}, journal={CMES - Computer Modeling in Engineering and Sciences}, author={Wu, Tsu-Yang and Meng, Qian and Yang, Lei and Kumari, Saru and Pirouz, Matin}, year={2022}, month=aug, pages={317–341} }

@article{X.Ren_J.Cao_B.Niu_L.Gan_Y.Zhang_L.Xiong_Y.Luo_H.Li_2024, title={A Formal Analysis of 5 G ProSe AKA Protocols for U2N Relay Communication}, ISSN={1941-0018}, DOI={10.1109/TDSC.2024.3522895}, journal={IEEE Transactions on Dependable and Secure Computing}, author={X. Ren and J. Cao and B. Niu and L. Gan and Y. Zhang and L. Xiong and Y. Luo and H. Li}, year={2024}, pages={1–15} }

@article{Xie_Liu_Ding_Tan_Han_2023, title={Provably Secure and Lightweight Patient Monitoring Protocol for Wireless Body Area Network in IoHT}, volume={2023}, url={https://www.scopus.com/inward/record.uri?eid=2-s2.0-85148772452&doi=10.1155%2f2023%2f4845850&partnerID=40&md5=b5dad521c2e0b3e85565ab1353e5cecb}, DOI={10.1155/2023/4845850}, journal={Journal of Healthcare Engineering}, author={Xie, Q. and Liu, D. and Ding, Z. and Tan, X. and Han, L.}, year={2023} }

@article{Y.Huang_G.Xu_X.Song_Y.Xu_2024, title={An Efficient RLWE-Based Privacy-Preserving Authentication Scheme Based on Edge Computing in Industrial Internet of Things}, volume={17}, ISSN={1939-1374}, DOI={10.1109/TSC.2024.3433534}, number={5}, journal={IEEE Transactions on Services Computing}, author={Y. Huang and G. Xu and X. Song and Y. Xu}, year={2024}, month=oct, pages={2012–2026} }

@inproceedings{Y.Song_F.Jiang_S.W.AliShah_R.Doss_2023, title={Multi-Factor Continuous Authentication of Drivers Leveraging Smartphone}, DOI={10.1109/SWC57546.2023.10449311}, note={journalAbbreviation: 2023 IEEE Smart World Congress (SWC)}, booktitle={2023 IEEE Smart World Congress (SWC)}, author={Y. Song and F. Jiang and S. W. Ali Shah and R. Doss}, year={2023}, month=aug, pages={1–9} }

@article{You_Kim_Pawana_Ko_2024, title={Mitigating Security Vulnerabilities in 6G Networks: A Comprehensive Analysis of the DMRN Protocol Using SVO Logic and ProVerif}, volume={14}, url={https://www.scopus.com/inward/record.uri?eid=2-s2.0-85208490035&doi=10.3390%2fapp14219726&partnerID=40&md5=2555ecef7e92fc8be674692c06c03dc6}, DOI={10.3390/app14219726}, number={21}, journal={Applied Sciences (Switzerland)}, author={You, I. and Kim, J. and Pawana, I.W.A.J. and Ko, Y.}, year={2024} }

@article{Zarbi_Zaeembashi_Bagheri_Adeli_2024, title={Toward designing a lightweight RFID authentication protocol for constrained environments}, volume={18}, DOI={10.1049/cmu2.12794}, number={14}, journal={IET Communications}, author={Zarbi, N. and Zaeembashi, A. and Bagheri, N. and Adeli, M.}, year={2024}, pages={846–859} }

@inbook{Zhu_Xu_Cui_2024, title={Formal Analysis of 5G EAP-TLS 1.3}, volume={193}, DOI={10.1007/978-3-031-53555-0_14}, booktitle={Lecture Notes on Data Engineering and Communications Technologies}, author={Zhu, N. and Xu, J. and Cui, B.}, year={2024}, pages={140–151} }

@article{Zou_Cao_Lu_Wang_Xu_Ma_Cheng_Xi_2024, title={A robust and effective 3-factor authentication protocol for smart factory in IIoT}, volume={220}, ISSN={0140-3664}, DOI={10.1016/j.comcom.2024.04.011}, abstractNote={Smart factory, as an intelligent application of industrial internet of things (IIoT), significantly enhances the efficiency of industrial processes while also reducing resource waste. In a system where security is crucial, the user authentication mechanism is essentially designed to prevent a range of security issues, including production failures caused by illegal intrusions from hackers. To verify the user’s identity and ensure data be accessed with a secure session key, a large number of authentication protocols have been provided for IIoT. However, existing newly alternatives show deficiency either in terms of superior performance or robust security. To get a better balance of security and efficiency, this paper designs a robust and effective 3-factor user authentication protocol. Then this paper shows the detailed security analyses and automated verification by the Proverif tool, where the ProVerif tool used in this paper can detect whether the eCK adversary with stronger attack ability can break the protocol, while the general ProVerif can only detect threat of the Dolev–Yao adversary with weaker attack ability to the protocol security. Subsequently, this paper presents performance comparison which indicates that the proposed protocol can be superior to those newly alternatives. Especially, the comparison results show that the computation cost consumed in the proposed protocol can be reduced by 62.2% than the average cost of all six compared alternatives. Lastly, the evaluation results on the energy consumption and the network delay indicate the practicability of proposed protocol for smart factory.}, journal={Computer Communications}, author={Zou, Shihong and Cao, Qiang and Lu, Ruichao and Wang, Chenyu and Xu, Guoai and Ma, Huanhuan and Cheng, Yingyi and Xi, Jinwen}, year={2024}, month=apr, pages={81–93} }

@inproceedings{Cortier_Grimm_Lallemand_Maffei_2017, address={Dallas Texas USA}, title={A Type System for Privacy Properties}, ISBN={978-1-4503-4946-8}, url={https://dl.acm.org/doi/10.1145/3133956.3133998}, DOI={10.1145/3133956.3133998}, abstractNote={Mature push button tools have emerged for checking trace properties (e.g. secrecy or authentication) of security protocols. The case of indistinguishability-based privacy properties (e.g. ballot privacy or anonymity) is more complex and constitutes an active research topic with several recent propositions of techniques and tools.}, booktitle={Proceedings of the 2017 ACM SIGSAC Conference on Computer and Communications Security}, publisher={ACM}, author={Cortier, Véronique and Grimm, Niklas and Lallemand, Joseph and Maffei, Matteo}, year={2017}, month=oct, pages={409–423}, language={en} }

@inbook{Focardi_Gorrieri_Martinelli_2004, address={Berlin, Heidelberg}, series={Lecture Notes in Computer Science}, title={Classification of Security Properties: (Part II: Network Security)}, volume={2946}, ISBN={978-3-540-20955-3}, url={http://link.springer.com/10.1007/978-3-540-24631-2\_4}, DOI={10.1007/978-3-540-24631-2\_4}, booktitle={Foundations of Security Analysis and Design II}, publisher={Springer Berlin Heidelberg}, author={Focardi, Riccardo and Gorrieri, Roberto and Martinelli, Fabio}, editor={Focardi, Riccardo and Gorrieri, Roberto}, year={2004}, pages={139–185}, collection={Lecture Notes in Computer Science}, language={en} }

@inproceedings{10.5555/647544.730459,
author = {Focardi, Riccardo and Martinelli, Fabio},
title = {A Uniform Approach for the Definition of Security Properties},
year = {1999},
isbn = {3540665870},
publisher = {Springer-Verlag},
address = {Berlin, Heidelberg},
booktitle = {Proceedings of the Wold Congress on Formal Methods in the Development of Computing Systems-Volume I - Volume I},
pages = {794–813},
numpages = {20},
series = {FM '99}
}

@inproceedings{Rouland_Hamid_Bodeveix_Filali_2019, address={Guangzhou, China}, title={A Formal Methods Approach to Security Requirements Specification and Verification}, rights={https://ieeexplore.ieee.org/Xplorehelp/downloads/license-information/IEEE.html}, ISBN={978-1-7281-4646-1}, url={https://ieeexplore.ieee.org/document/8882749/}, DOI={10.1109/ICECCS.2019.00033}, abstractNote={The speciﬁcation and the veriﬁcation of security requirements is one of the major computer-based systems challenges. Security requirements need to be precisely speciﬁed before a tool can manipulate them, and though several approaches to security requirements speciﬁcation have been proposed, they do not provide the scalability and ﬂexibility required in practice. We take this problem towards an integrated approach for security requirement speciﬁcation and treatment during the software architecture design time. The general idea of the approach is to: (1) specify security requirements as properties of a modeled system in a technology-independent speciﬁcation language; (2) implement the developed model in a suitable language with tool support for requirement satisfaction through model veriﬁcation; and (3) suggest a set of security policies to constrain the operation of the system and to guarantee the security properties. In the scope of this paper, we use ﬁrst-order logic as a formalism that is abstract and technology-independent and Alloy as a tooled language used in modeling and software development. To validate our work, we explore a set of representative security properties from categories based on CIA classiﬁcation in the context of secure component-based software architecture development.}, booktitle={2019 24th International Conference on Engineering of Complex Computer Systems (ICECCS)}, publisher={IEEE}, author={Rouland, Quentin and Hamid, Brahim and Bodeveix, Jean-Paul and Filali, Mamoun}, year={2019}, month=nov, pages={236–241}, language={en} }

@article{Sayar_Messe_Ebersold_Bruel_2025, title={From What to How: A Taxonomy of Formalized Security Properties}, url={http://arxiv.org/abs/2505.14514}, DOI={10.48550/arXiv.2505.14514}, abstractNote={Confidentiality, integrity, availability, authenticity, authorization, and accountability are known as security properties that secure systems should preserve. They are usually considered as security final goals that are achieved by system development activities, either in a direct or an indirect manner. However, these security properties are mainly elicited in the highlevel requirement phase during the System Development Life Cycle (SDLC) and are not refined throughout the latter phases as other artifacts such as attacks, defenses, and system assets. To align security properties refinement with attacks, defenses, and system assets refinements, we propose an SDLC taxonomy of security properties that may be used in a self-adaptive context and present the methodology for defining it. To verify and check the correctness of the resulting taxonomy, we use the Event-B formal language.}, note={arXiv:2505.14514 [cs]}, number={arXiv:2505.14514}, publisher={arXiv}, author={Sayar, Imen and Messe, Nan and Ebersold, Sophie and Bruel, Jean-Michel}, year={2025}, month=may, language={en} }

@article{Aldini_2006, title={Classification of security properties in a Linda-like process algebra}, volume={63}, rights={https://www.elsevier.com/tdm/userlicense/1.0/}, ISSN={01676423}, DOI={10.1016/j.scico.2005.07.010}, abstractNote={We provide a classiﬁcation of noninterference-based security properties for the formal analysis of secure information ﬂow in concurrent and distributed systems. This is done in the setting of a process algebra modeling some Linda coordination primitives (asynchronous communication and read operation). For this purpose, we deﬁne relaxed notions of behavioural equivalence that take into account the observational power of the external observer. The resulting taxonomy is compared with analogous security deﬁnitions based on synchronous communication models, thus emphasizing the inﬂuence of the Linda coordination model upon the expressivity of the security properties, by giving a new intuition to the relative merits.}, number={1}, journal={Science of Computer Programming}, author={Aldini, Alessandro}, year={2006}, month=nov, pages={16–38}, language={en} }

@book{10.5555/2141100,
author = {Jackson, Daniel},
title = {Software Abstractions: Logic, Language, and Analysis},
year = {2012},
isbn = {0262017156},
publisher = {The MIT Press},
address = {USA}
}

@book{10.5555/1855020,
author = {Abrial, Jean-Raymond},
title = {Modeling in Event-B: System and Software Engineering},
year = {2010},
isbn = {0521895561},
publisher = {Cambridge University Press},
address = {USA},
edition = {1st}
}

@article{hermann_taxonomy_2025,
    title = {A taxonomy of functional security features and how they can be located},
    volume = {30},
    issn = {1382-3256, 1573-7616},
    url = {https://link.springer.com/10.1007/s10664-025-10649-7},
    doi = {10.1007/s10664-025-10649-7},
    language = {en},
    number = {5},
    urldate = {2025-08-14},
    journal = {Empirical Software Engineering},
    author = {Hermann, Kevin and Schneider, Simon and Tony, Catherine and Yardim, Asli and Peldszus, Sven and Berger, Thorsten and Scandariato, Riccardo and Sasse, M. Angela and Naiakshina, Alena},
    month = oct,
    year = {2025},
    pages = {117},
}

@inproceedings{Belfaik_Lotfi_Sadqi_Safi_2024, address={Cham}, title={A Comparative Study of Protocols’ Security Verification Tools: Avispa, Scyther, ProVerif, and Tamarin}, ISBN={978-3-031-68653-5}, DOI={10.1007/978-3-031-68653-5_12}, abstractNote={The proliferation of Internet and network-based services has experienced significant growth, leading to the creation of novel security protocols across many different systems. This includes communication protocols, authentication protocols, group key management schemes, etc. Nevertheless, the complicated structure of these protocols frequently poses challenges for humans to evaluate and validate them properly. To ensure the efficiency of these security protocols, it is crucial to examine and verify them thoroughly before deployment. Consequently, researchers have developed user-friendly automatic validation tools that generate easily comprehensible output. In this paper, we performed a comparative examination of the four most commonly used security analysis and verification tools in the literature: AVISPA, Scyther, ProVerif, and Tamarin. Gaining a comprehensive understanding of each tool’s distinct attributes, functionalities, domains of application, and constraints will enable researchers and professionals to make well-informed decisions when selecting the most suitable tool for their requirements.}, booktitle={Digital Technologies and Applications}, publisher={Springer Nature Switzerland}, author={Belfaik, Yousra and Lotfi, Yousra and Sadqi, Yassine and Safi, Said}, editor={Motahhir, Saad and Bossoufi, Badre}, year={2024}, pages={118–128}, language={en} }

@Inbook{Krawczyk2019,
author="Krawczyk, Hugo",
editor="Jajodia, Sushil
and Samarati, Pierangela
and Yung, Moti",
title="Perfect Forward Secrecy",
bookTitle="Encyclopedia of Cryptography, Security and Privacy",
year="2019",
publisher="Springer Berlin Heidelberg",
address="Berlin, Heidelberg",
pages="1--3",
isbn="978-3-642-27739-9",
doi="10.1007/978-3-642-27739-9_90-2",
}

@manual{tamarinmanual,
  title        = {Tamarin Prover Manual: Security protocol analysis in the symbolic model},
  author       = {{The Tamarin Team}},
  year         = {2024},
  month        = {October},
  note         = {\url{https://tamarin-prover.com/manual/master/tex/tamarin-manual.pdf}},
  howpublished = {\url{https://tamarin-prover.com/manual/}},
}

@inproceedings{10.1145/1062455.1062491,
author = {Holmes, Reid and Murphy, Gail C.},
title = {Using structural context to recommend source code examples},
year = {2005},
isbn = {1581139632},
publisher = {Association for Computing Machinery},
address = {New York, NY, USA},
url = {https://doi.org/10.1145/1062455.1062491},
doi = {10.1145/1062455.1062491},
booktitle = {Proceedings of the 27th International Conference on Software Engineering},
pages = {117–125},
numpages = {9},
location = {St. Louis, MO, USA},
series = {ICSE '05}
}

@inproceedings{Zheng_1998, address={Berlin, Heidelberg}, title={Signcryption and its applications in efficient public key solutions}, ISBN={978-3-540-69767-1}, abstractNote={Signcryption is a new paradigm in public key cryptography that simultaneously fulfills both the functions of digital signature and public key encryption in a logically single step, and with a cost significantly lower than that required by the traditional “signature followed by encryption” approach. This paper summarizes currently known construction methods for signcryption, carries out a comprehensive comparison between signcryption and “signature followed by encryption”, and suggests a number of applications of signcryption in the search of efficient security solutions based on public key cryptography.}, booktitle={Information Security}, publisher={Springer Berlin Heidelberg}, author={Zheng, Yuliang}, editor={Okamoto, Eiji and Davida, George and Mambo, Masahiro}, year={1998}, pages={291–312} }

@InProceedings{10.1007/978-3-030-65277-7_8,
author="Kobeissi, Nadim
and Nicolas, Georgio
and Tiwari, Mukesh",
editor="Bhargavan, Karthikeyan
and Oswald, Elisabeth
and Prabhakaran, Manoj",
title="Verifpal: Cryptographic Protocol Analysis for the Real World",
booktitle="Progress in Cryptology --  INDOCRYPT 2020",
year="2020",
publisher="Springer International Publishing",
address="Cham",
pages="151--202",
isbn="978-3-030-65277-7"
}

@article{dolev_security_1983,
	title = {On the security of public key protocols},
	volume = {29},
	issn = {1557-9654},
	url = {https://ieeexplore.ieee.org/document/1056650},
	doi = {10.1109/TIT.1983.1056650},
	number = {2},
	urldate = {2024-06-10},
	journal = {IEEE Transactions on Information Theory},
	author = {Dolev, D. and Yao, A.},
	month = mar,
	year = {1983},
	note = {Conference Name: IEEE Transactions on Information Theory},
	pages = {198--208},
}

@article{10.1145/2363.2433,
author = {Gelernter, David},
title = {Generative communication in Linda},
year = {1985},
issue_date = {Jan. 1985},
publisher = {Association for Computing Machinery},
address = {New York, NY, USA},
volume = {7},
number = {1},
issn = {0164-0925},
url = {https://doi.org/10.1145/2363.2433},
doi = {10.1145/2363.2433},
journal = {ACM Trans. Program. Lang. Syst.},
month = jan,
pages = {80–112},
numpages = {33}
}

@inproceedings{10.1145/3372297.3423354,
author = {Cremers, Cas and Fairoze, Jaiden and Kiesl, Benjamin and Naska, Aurora},
title = {Clone Detection in Secure Messaging: Improving Post-Compromise Security in Practice},
year = {2020},
isbn = {9781450370899},
publisher = {Association for Computing Machinery},
address = {New York, NY, USA},
url = {https://doi.org/10.1145/3372297.3423354},
doi = {10.1145/3372297.3423354},
booktitle = {Proceedings of the 2020 ACM SIGSAC Conference on Computer and Communications Security},
pages = {1481–1495},
numpages = {15},
location = {Virtual Event, USA},
series = {CCS '20}
}

@inproceedings{292d45b0bcaf4dd3a75b2b2ba47bd853,
title = "Towards Secure IoT Deployments: A DSL and Digital Twin-Based Emulation Platform for Security Verification",
author = "Leonard Tudorache",
year = "2025",
month = dec,
day = "10",
doi = "10.1109/MODELS-C68889.2025.00017",
language = "English",
pages = "71--76",
booktitle = "2025 ACM/IEEE 28th International Conference on Model Driven Engineering Languages and Systems Companion, MODELS-C 2025",
publisher = "Institute of Electrical and Electronics Engineers",
address = "United States",
note = "28th ACM/ IEEE International Conference on Model Driven Engineering Languages and Systems Companion, MODELS-C 2025, MODELS-C 2025 ; Conference date: 05-10-2025 Through 10-10-2025",
}

@software{Basin_Tamarin_Prover,
author = {Basin, David and Cremers, Cas and Dreier, Jannik and Meier, Simon and Sasse, Ralf and Schmidt, Benedikt},
title = {{Tamarin Prover}},
url = {https://github.com/tamarin-prover/tamarin-prover}
}

@article{tawalbeh_iot_2020,
    title = {{IoT} {Privacy} and {Security}: {Challenges} and {Solutions}},
    volume = {10},
    copyright = {https://creativecommons.org/licenses/by/4.0/},
    issn = {2076-3417},
    shorttitle = {{IoT} {Privacy} and {Security}},
    url = {https://www.mdpi.com/2076-3417/10/12/4102},
    doi = {10.3390/app10124102},
    language = {en},
    number = {12},
    urldate = {2025-02-19},
    journal = {Applied Sciences},
    author = {Tawalbeh, Lo’ai and Muheidat, Fadi and Tawalbeh, Mais and Quwaider, Muhannad},
    month = jun,
    year = {2020},
    pages = {4102},
}

@incollection{diffie2022new,
  title={New directions in cryptography},
  author={Diffie, Whitfield and Hellman, Martin E},
  booktitle={Democratizing cryptography: the work of Whitfield Diffie and Martin Hellman},
  pages={365--390},
  year={2022}
}

\end{document}